\documentclass[11pt]{article}
\pdfoutput=1

\usepackage[dvipsnames]{xcolor}
\usepackage[a4paper]{geometry}
\usepackage[english]{babel}
\usepackage{cite,enumerate,enumitem,booktabs,float,graphicx}

\usepackage[T1]{fontenc}
\usepackage[utf8]{inputenc}
\usepackage{lmodern}

\makeatletter
\g@addto@macro\bfseries{\boldmath}
\makeatother

\usepackage{bm,amsmath,amssymb,tensor,mathtools}
\numberwithin{equation}{section}

\usepackage{scalerel}
\newcommand{\osb}[2]{\overset{\makebox[0pt]{$\scriptscriptstyle{\scaleto{(\hspace{-0.06em}#2\hspace{-0.06em})}{4.5pt}}$}}{#1}{}}

\newcommand{\nablaCc}{\osb{\nabla}{\check{C}}}
\newcommand{\nablaCt}{\osb{\nabla}{\tilde{C}}}
\newcommand{\nablaC}{\osb{\nabla}{C}}

\usepackage{tikz}
\usetikzlibrary{cd}

\usepackage{caption}
\usepackage{subcaption}

\usepackage[pdftex]{hyperref}
\definecolor{dark-blue}{rgb}{0.15,0.15,0.4}
\hypersetup{
  colorlinks,
  linkcolor={dark-blue},
  citecolor={blue}
}

\newcommand{\pd}{\partial}

\newcommand{\LL}{\mathcal{L}}

\newcommand{\inv}{^{-1}}

\newcommand{\mK}{\mathcal{K}}

\usepackage{authblk}

\title{\textbf{%
  Setting the connection free
  \\
  in the Galilei and Carroll expansions of gravity
  }
}
\date{}

\author[1]{Jørgen Musaeus}
\author[2,3]{Niels A.\ Obers}
\author[1]{Gerben Oling}
\affil[1]{%
  School of Mathematics and Maxwell Institute for Mathematical Sciences,
  \protect\\
  University of Edinburgh,
  Peter Guthrie Tait Road,
  Edinburgh EH9 3FD, UK
}
\affil[2]{%
  Nordita,
  KTH Royal Institute of Technology and Stockholm University,
  \protect\\
  Hannes Alfv\'ens v\"ag 12, SE-106 91 Stockholm, Sweden
}
\affil[3]{%
  Niels Bohr International Academy, The Niels Bohr Institute,
  University of Copenhagen,
  \protect\\
  Blegdamsvej 17,
  DK-2100 Copenhagen Ø,
  Denmark
}

\begin{document}
\maketitle
\thispagestyle{empty}

\begin{abstract}

We obtain a Palatini-type formulation for the Galilei and Carroll expansions of general relativity, where the connection is promoted to a variable. 
Known versions of these large and small speed of light expansions are derived from the Einstein--Hilbert action and involve dynamical Newton--Cartan or Carroll geometry,
along with additional gauge fields at subleading orders.
The corresponding Palatini actions that we obtain in this paper are derived from an appropriate expansion of the Einstein--Palatini action, and the connection variable reduces to the Galilei- or Carroll-adapted connection on shell.
In particular, we present the Palatini form for the next-to-leading order Galilean action
and recover the known equations of motion.
We also compute the leading-order Palatini-type action for the Carrollian case and show that,
while it depends on the connection variable,
it reduces on shell to the known action of electric Carroll gravity,
which only depends on extrinsic curvature.
\end{abstract}

\newpage
\tableofcontents

\section{Introduction}%

Recent years have witnessed a resurgence of interest in both the  large and small speed of light expansion of general relativity, which lead to non-relativistic or Galilean and ultra-local or Carroll approximations, respectively. 
These developments have been  fueled in part by a deeper understanding and revival \cite{Andringa:2010it,Christensen:2013lma,Hartong:2015zia}  of non-Lorentzian geometry and its diverse connections to field theory \cite{Son:2013rqa,Jensen:2014aia,Hartong:2014oma,Geracie:2014nka}, holography  and string theory \cite{Andringa:2012uz,Harmark:2017rpg,Bergshoeff:2018yvt}.%
\footnote{For more references and background see the recent reviews  on 
non-relativistic gravity  \cite{Hartong:2022lsy}, non-Lorentzian geometry \cite{Bergshoeff:2022eog}, non-relativistic string theory  \cite{Oling:2022fft}
and non-relativistic field theory~\cite{Baiguera:2023fus}.} 
In particular, a covariant, off-shell large speed of light expansion of general relativity was obtained in 
\cite{VandenBleeken:2017rij,Hansen:2019pkl,Hansen:2020pqs} describing, in principle, the dynamics of gravity
at any particular order in an expansion of $1/c^2$, with $c$ the speed of light.  
This expansion
leads to a novel `type II' Newton--Cartan geometry,
which turns out to be essential to construct an action for Newtonian gravity in terms of dynamical geometry.
The actions obtained from the large speed of light expansion also go beyond Newtonian gravity, since we are not necessarily performing a weak-field expansion.
Furthermore, a key observation has been to emphasize the role of torsion in Newton-Cartan 
geometry.
While this quantity is zero for Newtonian physics, corresponding to absolute time, it does not need to be zero in general, and thus torsional Newton--Cartan geometry can describe gravitational time dilation even in the non-relativistic regime. 
Using similar methods, the corresponding Carrollian or ultra-local expansion for small speed of light was developed in \cite{Hansen:2021fxi}, making contact with earlier work on Carroll limits and expansions of gravity in~\cite{Henneaux:1979vn,Dautcourt:1997hb,Bergshoeff:2017btm,Henneaux:2021yzg}.

There are by now ample reasons to consider such expansions, of which we mention here just a few.
For the Galilean case, they include covariant formulations of post-Newtonian physics \cite{Tichy:2011te,Hartong:2023ckn,pn-nc-companionpaper},
understanding the coupling of  non-relativistic quantum matter to geometry \cite{Hartong:2023yxo},
the relation to low-energy effective actions in non-relativistic string theory \cite{Gomis:2019zyu,Gallegos:2019icg,Bergshoeff:2019pij,Bergshoeff:2021bmc},
non-Lorentzian supergravity  (see the review \cite{Bergshoeff:2022iyb})
and numerous additional ones summarized in the review \cite{Hartong:2022lsy}.
Likewise, for Carrollian gravity there are interesting
relations to cosmology \cite{deBoer:2021jej},
applications to black holes \cite{Penna:2018gfx,Donnay:2019jiz,Hansen:2021fxi,Perez:2022jpr,deBoer:2023fnj,Ecker:2023uwm} 
and suggestive simplifications of the 3+1 formulation of general relativity \cite{Hansen:2021fxi}
(see Refs.~\cite{Duval:2014uva,Hartong:2015xda,Bekaert:2015xua,Ciambelli:2019lap,Niedermaier:2020jdy,Figueroa-OFarrill:2020gpr,Grumiller:2020elf,Gomis:2020wxp,Perez:2021abf,%
  Figueroa-OFarrill:2021sxz,Herfray:2021qmp,Figueroa-OFarrill:2022mcy,Baiguera:2022lsw,Fuentealba:2022gdx,Campoleoni:2022ebj,Campoleoni:2022wmf,Bergshoeff:2023rkk,Ciambelli:2023tzb,%
Blair:2023noj,Bergshoeff:2023vfd,Ekiz:2022wbi} for further related aspects of Carroll gravity and geometry).

In principle, the work of~\cite{Hansen:2019pkl,Hansen:2020pqs,Hansen:2021fxi}
allows us to obtain both the non-relativistic and ultra-local expansions of the action and equations of motions (EOMs) of general relativity (GR) to any order.
Furthermore, there is an understanding of the underlying symmetry principle at any given order, based on a  truncation of an appropriate Lie algebra expansion of the Poincar\'e algebra~\cite{Hansen:2019pkl,Bergshoeff:2019ctr,Gomis:2019sqv,Hansen:2020pqs}.

However, developing this geometric expansion in the speed of light becomes increasingly challenging at higher orders.
So far, the study of the action and EOMs has primarily been done in terms of what one could call a second-order formalism for the geometrical variables, using non-Lorentzian geometric fields that are the analogue of the metric in Lorentzian geometry and a particular connection compatible with this structure.
In this language, explicit expressions for the complete expanded actions only exist up to next-to-next-to-leading order in the non-relativistic case~\cite{Hansen:2020pqs} and next-to-leading order in the ultra-local case~\cite{Hansen:2021fxi}.
On the other hand, in GR we know that several computations simplify somewhat in a first-order or Palatini formulation, where the connection is promoted to a dynamical variable which reproduces the Levi-Civita connection on shell.

A natural question is therefore whether there exists an alternate, and perhaps more efficient,
Palatini-type formulation for the Galilean and Carrollian expansions of GR,%
\footnote{For the case of Einstein gravity, the Palatini action is in fact also a first order action. In general, however, the Palatini actions we will develop in this work (by treating the metric and connection as independent degrees of freedom) are not necessarily fully first-order in derivatives on the metric due to terms related to the torsion.
Using the term `first-order' actions in this context can thus be a slight abuse of language, and we will mainly avoid it in favor of `Palatini' or `Palatini-type' actions.}
and some work on this has already been done~\cite{Bergshoeff:2017btm,Cariglia:2018hyr,Guerrieri:2020vhp,Hansen:2020wqw,Guerrieri:2021cdz}.
In particular, in \cite{Hansen:2020wqw} a novel Palatini-type formulation of GR was obtained where the connection EOM reproduces the Galilean-compatible connection suitable for the non-relativistic expansion.
In the process, this work elucidated and emphasized the appearance of torsion in the non-relativistic expansion of the connection, which was missed in early work on Carroll and Galilei limits~\cite{Bergshoeff:2017btm}
(but see~\cite{Bergshoeff:2023rkk} for recent work on limits incorporating torsion).

In this paper, we will take the next step and obtain a Palatini-type formulation relevant to both the non-relativistic as well as ultra-local expansion of GR.
We will thus obtain Palatini actions for the Galilei and Carroll expansions of gravity. 
Building on~\cite{Hansen:2020wqw} we explicitly show that we can recover the appropriate connections from the equations of motions.
Our main motivation is the expected simplification of the computation of EOMs at higher order,
and we therefore also demonstrate that our novel Palatini actions allow us to recover known second-order metric equations of motion after putting the connection on shell.
In addition, we expect this formulation to aid in the study of various physical applications, including the construction of boundary charges,
as mentioned in the Discussion.

Relatedly, since constructing further subleading actions is still cumbersome using our Palatini-type actions,
one may wonder whether such actions can instead be obtained by an algebraic procedure,
using the known symmetry principle mentioned above.
This was considered in several works, especially for Carroll limits, but these approaches often fail to reproduce what we would call the leading-order (or `electric') Carroll action~\cite{Bergshoeff:2017btm,Figueroa-OFarrill:2022mcy,Campoleoni:2022ebj}, which again is related to the importance of torsion (see also~\cite{Bergshoeff:2023rkk}).
Since this is relatively novel territory, the explicit forms of the Palatini-type actions
for next-to-leading order Galilean gravity
and leading-order Carroll gravity
are by themselves a useful milestone, and comprise one of our main results.
Perhaps surprisingly, we will find that the leading-order Palatini action for Carroll gravity depends on the connection, even though its known second-order form only depends on the extrinsic curvature and is hence independent of the connection. 

A brief outline and summary of the paper is as follows. In Section \ref{sec:ep-review}
we review the Palatini description of Einstein gravity.
Treating the connection as an independent variable, the solution of its equation of motion gives the metric-compatible torsion-free Levi-Civita connection, up to a one-form ambiguity.
We then show that there is a slightly different approach in which one shifts the connection degree of freedom using the Levi-Civita connection, which will be used later on. 
In Section \ref{sec:pal-action} we first review the pre-non-relativistic (PNR) and pre-ultra-local (PUL) decomposition of  GR  following \cite{Hansen:2020pqs,Hansen:2021fxi}.
This is essentially a rewriting for GR in which time and space are split
covariantly, in anticipation of the large/small speed of light expansion.
It involves a decomposition of the metric
separating time-like and space-like directions along with an appropriate decomposition of the Levi-Civita
connection, the details of which depends on whether one considers large or small speed of light. 
This procedure is what defines for us the PNR/PUL connection, adapted to the Galilean or Carrollian structures that emerge in the large/small speed of light expansions.
We can use this to mimic the Palatini procedure for the PNR/PUL
actions, which is what we obtain next in Section~\ref{sec:exp}.
We show that one can solve the PNR and PUL connection from its equation of motion in these actions.
For the Galilean case, we will then write down the resulting leading-order (LO) and next-to-leading order (NLO) actions, while for the Carroll case 
we will confine ourselves to presenting the leading-order action. 
Then in Section \ref{sec:metric-eom} we put our Palatini formulation to the test by computing the metric equations of motion
from the NLO Galilean action using the action obtained in the previous section. 
We end in Section~\ref{sec:discussion} with an outlook.
In Appendix \ref{app:geom-id} we list conventions and geometric identities that will be used in the main text.
Finally, Appendix \ref{app:solving-connection-for-ep} gives further
computational details of the solution of the general connection equations of motion in the Einstein–Palatini action.

\section{Einstein-Palatini review}
\label{sec:ep-review}
We first review the description of general relativity using the Einstein--Palatini action,
\begin{equation}
  \label{eq:einstein-palatini-action}
  S_\text{EP}[g,\Gamma]
  = \frac{1}{2\kappa} \int d^{d}x \sqrt{-g}\,
  g^{\mu\nu} R_{\mu\nu}\,.
\end{equation}
This action can be obtained from the Einstein--Hilbert action by promoting the Levi-Civita connection to an independent variable $\Gamma^\rho_{\mu\nu}$,
so that in particular the Ricci tensor $R_{\mu\nu}$ depends on the connection but not on the metric.
Sometimes the connection variable is assumed to be symmetric, so that its torsion always vanishes.
We will not impose such a restriction in order to make the comparison with our non-Lorentzian discussion below more obvious, as in those cases torsion is generically nonzero.

Using our definitions and conventions in Appendix~\ref{app:geom-id}, and in particular the variation of a general Ricci tensor obtained from Equation~\eqref{eq:general-riemann-tensor-conn-var}, the variation of this action is
\begin{equation}
  \label{eq:einstein-palatini-action-variation}
  \begin{split}
    \delta S_\text{EP}
    &= \frac{1}{2\kappa} \int d^dx \sqrt{-g}\,
    \left(R_{\mu\nu} - \frac{1}{2}g_{\mu\nu} R\right) \delta g^{\mu\nu}
    \\
    &{}\qquad
    + \frac{1}{2\kappa} \int d^dx \sqrt{-g}\,
    g^{\mu\nu} \left(
      - \nabla_\mu \delta\Gamma^\rho_{\rho\nu}
      + \nabla_\rho \delta\Gamma^\rho_{\mu\nu}
      - T^\lambda{}_{\mu\rho} \delta\Gamma^\rho_{\lambda\nu}
    \right).
  \end{split}
\end{equation}
As we will discuss below, the resulting connection equations of motion can be solved to obtain the Levi-Civita connection.
With the connection on shell, the metric equations of motion then reproduce the Einstein equations.
However, before we show how to solve the connection equation of motion, let us emphasize a few key points.

First, as we already mentioned, the Einstein--Palatini action has exactly the same form as the usual Einstein--Hilbert action, with the only difference between the two being that the connection is a variable in the former.
However, this is not generic.
Promoting the connection to a variable in for example a higher-derivative theory of gravity typically does not lead to connection equations of motion that can be solved to obtain the Levi-Civita connection and to recover the original action~\cite{Borunda:2008kf}.
In the following, we will introduce reparametrizations of general relativity using variables that are adapted to the non-relativistic or ultra-local expansions.
As we will see, the corresponding reparametrizations of the Einstein--Palatini action are not just obtained by allowing the connection in the reparametrization of the Einstein--Hilbert action to vary, but they contain additional terms.
In total, this allows us to solve the actions for the corresponding connection and reproduce the correct actions with the connection on shell.

Relatedly, the metric equations of motion coming from the Einstein--Palatini action in~\eqref{eq:einstein-palatini-action-variation} are already of the same form as the metric equations of motion of the Einstein--Hilbert action.
All that remains to recover the Einstein equations is specifying the connection used in the Ricci tensor.
This is related to the fact that the variation of the Ricci tensor produces a boundary term when the connection is on shell, which will no longer be the case in our Galilei-adapted and Carroll-adapted variables below.
Instead, the terms coming from varying the Ricci tensor are reproduced by the variation of the additional terms in the corresponding reparametrizations of the Einstein--Palatini action.

\subsection{Solving for the connection directly}
\label{ssec:ep-review-solve-connection-direct}
Now let us return to obtaining and solving the connection equation of motion from the variation~\eqref{eq:einstein-palatini-action-variation} of the standard formulation of the Einstein--Palatini action.
We first rewrite the terms involving a derivative of the connection variation using a total covariant derivative, which gives
\begin{align}
  \label{eq:einstein-palatini-action-gamma-variation-rewrite}
  \delta_\Gamma S_\text{EP}
  &= \frac{1}{2\kappa} \int d^dx \Big[
    - \nabla_\alpha \left(
      \sqrt{-g}\, g^{\alpha\nu} \delta^\mu_\rho \delta\Gamma^\rho_{\mu\nu}
    \right)
    + \nabla_\rho \left(
      \sqrt{-g}\, g^{\mu\nu} \delta \Gamma^\rho_{\mu\nu}
    \right)
    \\
    &{}\qquad\qquad\qquad
    \nonumber
    + \delta\Gamma^\rho_{\mu\nu} \left(
      \delta^\mu_\rho \nabla_\lambda \left[\sqrt{-g}\, g^{\lambda\nu} \right]
      - \nabla_\rho \left[\sqrt{-g}\, g^{\mu\nu}\right]
      - \sqrt{-g}\, g^{\sigma\nu} T^\mu{}_{\sigma\rho}
    \right)
  \Big]\,.
\end{align}
The terms on the first line involve a total covariant derivative of a weight one vector density.
We can rewrite these using Equation~\eqref{eq:weight-one-vector-density-divergence},
which leads to
\begin{align}
  \label{eq:einstein-palatini-action-gamma-variation-rewrite-two}
  \delta_\Gamma S_\text{EP}
  &\approx \frac{1}{2\kappa} \int d^dx \Big[
    \delta\Gamma^\rho_{\mu\nu} \left(
      - \sqrt{-g}\, T^\beta{}_{\beta\alpha} g^{\alpha\nu} \delta^\mu_\rho
      + \sqrt{-g}\, T^\alpha{}_{\alpha\rho} g^{\mu\nu}
    \right)
    \\\nonumber
    &{}\qquad\qquad\qquad
    + \delta\Gamma^\rho_{\mu\nu} \left(
      \delta^\mu_\rho \nabla_\lambda \left[\sqrt{-g}\, g^{\lambda\nu} \right]
      - \nabla_\rho \left[\sqrt{-g}\, g^{\mu\nu}\right]
      - \sqrt{-g}\, T^\mu{}_{\alpha\rho} g^{\alpha\nu}
    \right)
  \Big]\,.
\end{align}
Here and in the following, we use `$\approx$' to denote that two quantities are equal up to boundary terms.
From this, we obtain the following connection equation of motion,
\begin{equation}
  \label{eq:einstein-palatini-direct-gamma-eom}
  \begin{split}
    0
    &=
    \delta^\mu_\rho \nabla_\alpha \left(\sqrt{-g}\, g^{\alpha\nu} \right)
    - \nabla_\rho \left(\sqrt{-g}\, g^{\mu\nu}\right)
    \\
    &{}\qquad
    - \sqrt{-g}\, T^\alpha{}_{\alpha\beta} g^{\beta\nu} \delta_\rho^\mu
    + \sqrt{-g}\, T^\alpha{}_{\alpha\rho} g^{\mu\nu}
    - \sqrt{-g}\, T^\mu{}_{\alpha\rho} g^{\alpha\nu}\,.
  \end{split}
\end{equation}
This can be interpreted as an equation for the `metricity density'
$\nabla_{\rho} \left(\sqrt{-g}\, g^{\mu\nu}\right)$
and the torsion tensor
$T^\rho{}_{\mu\nu}$.
As detailed in Appendix~\ref{app:solving-connection-for-ep}, it determines the connection to be
\begin{equation}
  \label{eq:einstein-palatini-direct-gamma-sol}
  \Gamma^\rho_{\mu\nu}
  = \osb{\Gamma}{LC}^\rho_{\mu\nu} + \delta^\rho_\nu A_\mu\,.
\end{equation}
Here, the first term is the usual Levi--Civita connection,
\begin{equation}
  \label{eq:levi-civita-connection}
  \osb{\Gamma}{LC}^\rho_{\mu\nu}
  = \frac{g^{\rho\lambda}}{2} \left(
    \pd_\mu g_{\nu\rho} +  \pd_\nu g_{\rho\mu} - \pd_\rho g_{\mu\nu}
  \right)\,,
\end{equation}
and the second term is an ambiguity~\cite{Julia:1998ys,Dadhich:2012htv,Bernal:2016lhq}
parameterized by a one-form $A_\mu$, which however drops out of the action~\eqref{eq:einstein-palatini-action} entirely.
One can exclude this ambiguity by assuming zero torsion from the start, as in for example~\cite{Misner:1973prb}.
Although they may in principle affect the coupling to matter, we will ignore such ambiguities that drop out of the action for our present purposes.%
\footnote{\label{fn:connection-gauge-symmetry}%
  Note that it has been argued in~\cite{Dadhich:2012htv,Bernal:2016lhq} that the connection ambiguity~\eqref{eq:einstein-palatini-direct-gamma-sol} does not affect the Einstein equation, nor does it physically affect the point particle coupling.
  It would be very interesting to see if similar results hold for the additional connection gauge symmetries that we obtain in the Galilean and Carrollian expansions below.
}

\subsection{Solving for the connection using a shift}
\label{ssec:ep-review-solve-connection-shift}
While the connection equation of motion obtained from the variation of the action~\eqref{eq:einstein-palatini-action} can be solved directly in that form, it is easier to take a slightly different approach.
Instead of varying with respect to $\Gamma^\rho_{\mu\nu}$, we introduce a change of variables
\begin{equation}
  \label{eq:connection-lc-shift}
  \Gamma^\rho_{\mu\nu}
  = \osb{\Gamma}{LC}^\rho_{\mu\nu}
  + S^\rho{}_{\mu\nu}\,,
\end{equation}
in terms of the Levi-Civita connection~\eqref{eq:levi-civita-connection}
and an arbitrary tensor $S^\rho{}_{\mu\nu}$.
Of course, this change of variables contains some hindsight, but since $S^\rho{}_{\mu\nu}$ is general it is without loss of generality.
Using Equation~\eqref{eq:general-riemann-tensor-conn-shift}, we can then rewrite the action~\eqref{eq:einstein-palatini-action} as
\begin{align}
  \label{eq:einstein-palatini-shift-action-prelim}
  S_\text{EP-S}[g,S]
  &= S_\text{EP}[g,\Gamma=\osb{\Gamma}{LC}+S]
  \\
  \nonumber
  &= \frac{1}{2\kappa} \int d^dx \sqrt{-g}\, g^{\mu\nu} \left[
    \osb{R}{LC}_{\mu\nu}
    - \osb{\nabla}{LC}_\mu S^\rho{}_{\rho\nu}
    + \osb{\nabla}{LC}_\rho S^\rho{}_{\mu\nu}
    - S^\rho{}_{\mu\lambda} S^\lambda{}_{\rho\nu}
    + S^\rho{}_{\rho\lambda} S^\lambda{}_{\mu\nu}
  \right]
  \\
  &\approx
  \label{eq:einstein-palatini-shift-action}
  S_\text{EH}[g]
  +
  \frac{1}{2\kappa} \int d^dx \sqrt{-g}\, g^{\mu\nu} \left(
    - S^\rho{}_{\mu\lambda} S^\lambda{}_{\rho\nu}
    + S^\rho{}_{\rho\lambda} S^\lambda{}_{\mu\nu}
  \right),
\end{align}
where we have dropped two total derivatives.
The first term on the last line is just the Einstein--Hilbert action.
The freedom in the connection variable, which is now parametrized by $S^\rho{}_{\mu\nu}$ through~\eqref{eq:connection-lc-shift}, is thus entirely captured by the last term, which is quadratic and tensorial.
Varying it gives rise to the equation of motion
\begin{equation}
  \label{eq:einstein-palatini-shift-eom}
  0 =
  - g^{\mu\lambda} S^\nu{}_{\rho\lambda}
  - g^{\nu\lambda} S^\mu{}_{\lambda\rho}
  + \delta^\mu_\rho S^\nu{}_{\alpha\beta} g^{\alpha\beta}
  + g^{\mu\nu} S^\lambda{}_{\lambda\rho}\,.
\end{equation}
As detailed in Appendix~\ref{app:solving-connection-for-ep},
we can solve this to obtain
\begin{equation}
  \label{eq:einstein-palatini-shift-s-sol}
  S^\rho{}_{\mu\nu}
  = \delta^\rho_\nu A_\mu\,.
\end{equation}
Together with~\eqref{eq:connection-lc-shift}, this reproduces precisely the solution~\eqref{eq:einstein-palatini-direct-gamma-sol} above.
Again, the ambiguity $A_\mu$ drops out of the action~\eqref{eq:einstein-palatini-shift-action}, and we recover the Einstein--Hilbert action upon putting the connection on shell.

Obtaining and solving the connection equation of motion is more straightforward in this approach, and we will see below that it carries over nicely to the Galilei and Carroll reparametrizations and expansions.
However, the variation of the action~\eqref{eq:einstein-palatini-shift-action} with respect to the metric is equivalent to the metric variation of the Einstein--Hilbert action.
As such, this reparametrization of the Palatini action does not provide any potential benefit for determining the metric equations of motion, in contrast to the original form of the action in~\eqref{eq:einstein-palatini-action} where the Ricci tensor was independent of the connection.
For the Galilei and Carroll reparametrizations and expansions of the Einstein--Palatini action,
we will therefore start from~\eqref{eq:einstein-palatini-action}, and we derive the metric equations of motion from the resulting actions in that form.
Meanwhile, for the connection equations of motion we will perform a change of variables such that we end up with the Galilei or Carroll equivalent of the form of the Palatini action in~\eqref{eq:einstein-palatini-shift-action}.

\section{Pre-non-relativistic and pre-ultra-local Palatini action}
\label{sec:pal-action}
As briefly mentioned in the Introduction, the Galilean and Carrollian expansion of general relativity proceeds in two steps.
First, the Einstein--Hilbert action is rewritten using variables adapted to the non-relativistic and ultra-local symmetries that will emerge in the corresponding expansion.
This change of variables is known as the `pre-non-relativistic' (PNR) \cite{Hansen:2019pkl,Hansen:2020pqs} and `pre-ultra-local' (PUL) \cite{Hansen:2021fxi} parameterization, respectively.
At this point, the resulting PNR and PUL parametrizations of the Einstein--Hilbert action are still equivalent to the Einstein--Hilbert action, only written in different variables.
We will generalize these reparametrizations to the Einstein--Palatini action in the present section.
The next section will then describe the non-relativistic and ultra-local Galilei and Carroll expansion using the adapted PNR and PUL variables.

As a first step in both parametrizations, we rewrite the Lorentzian metric as
\begin{equation}
  \label{eq:pnr-metric-decomposition}
  g_{\mu\nu}
  = - c^2 T_\mu T_\nu + \Pi_{\mu\nu}\,,
  \qquad
  g^{\mu\nu}
  = - \frac{1}{c^2} V^\mu V^\nu + \Pi^{\mu\nu}\,.
\end{equation}
The $T_\mu$ and $V^\mu$ are timelike vielbeine, while the $\Pi_{\mu\nu}$ and $\Pi^{\mu\nu}$ act roughly as spatial metrics.
These variables are assumed to be analytic in $1/c^2$ or $c^2$, and they satisfy the following orthonormality relations,
\begin{equation}
  \label{eq:pnr-on-completeness}
  V^\mu T_\mu = -1\,,
  \qquad
  V^\mu \Pi_{\mu\nu} = 0\,,
  \qquad
  T_\mu \Pi^{\mu\nu} = 0\,,
  \qquad
  \delta^\mu_\nu = - V^\mu T_\nu + \Pi^{\mu\rho} \Pi_{\rho\nu}\,.
\end{equation}
In addition to general diffeomorphisms, these variables transform under local Lorentz boosts, which will become local Galilei or Carroll boosts plus subleading corrections in the respective expansions.
Finally, the integration measure becomes
$\sqrt{-g} = c\,E$
where we have $E^2= - \det(- T_\mu T_\nu + \Pi_{\mu\nu})$.

Previously, this decomposition was applied to the Einstein--Hilbert action.
Depending on how we interpret the resulting decomposition of the Levi-Civita connection, we can obtain a connection that is adapted to either the Galilei or the Carroll structure that emerges in the $c\to\infty$ or $c\to0$ expansion, respectively.
We will first treat the PNR case with emerging Galilei structure in Sections~\ref{ssec:pal-action-pnr} and~\ref{ssec:pal-action-pnr-solve-conn}.
The PUL case with emerging Carroll structure then proceeds similarly and is discussed in Section~\ref{ssec:pal-action-pul}.

\subsection{Pre-non-relativistic decomposition of the Palatini action}
\label{ssec:pal-action-pnr}
In the pre-non-relativistic (PNR) case, the Levi-Civita connection is decomposed as
\begin{align}
  \label{eq:pnr-lc-decomposition}
  \osb{\Gamma}{LC}^\rho_{\mu\nu}
  &= c^2\, \osb{C}{-2}^\rho{}_{\mu\nu}
  + \check{C}^\rho_{\mu\nu} + \osb{C}{0}^\rho{}_{\mu\nu}
  + \frac{1}{c^2} \osb{C}{2}^\rho_{\mu\nu}\,,
\end{align}
Here, we have collected powers of $c$ that appear due to the metric decomposition~\eqref{eq:pnr-metric-decomposition}.
Additionally, we obtain from the $c^0$ terms a connection $\check{C}^\rho_{\mu\nu}$ that is adapted to the Galilean structure that emerges at leading order in the $c\to\infty$ expansion,
\begin{equation}
  \label{eq:pnr-connection}
  \check{C}^\rho_{\mu\nu}
  = - V^\rho \pd_\mu T_\nu
  + \frac{1}{2} \Pi^{\rho\sigma} \left(
    \pd_\mu \Pi_{\nu\sigma} + \pd_\nu \Pi_{\sigma\mu} - \pd_\sigma \Pi_{\mu\nu}
  \right).
\end{equation}
This means in particular that the corresponding covariant derivative $\nablaCc_\rho$ satisfies
\begin{equation}
  \label{eq:pnr-conn-met-comp-a}
  \nablaCc_\rho T_\mu
  = 0 \,,
  \qquad
  \nablaCc_\rho \Pi^{\mu\nu}
  = 0 \,.
\end{equation}
Additionally, we have
\begin{equation}
  \label{eq:pnr-conn-met-comp-b-torsion}
  \nablaCc_\rho V^\mu
  = - \Pi^{\mu\nu} \mK_{\nu\rho}\,,
  \qquad
  \nablaCc_\rho \Pi_{\mu\nu}
  = -2 T_{(\mu} \mK_{\nu)\rho}\,,
  \qquad
  \osb{T}{\check{C}}^\rho{}_{\mu\nu}
  = 2 \check{C}^\rho_{[\mu\nu]}
  = - V^\rho T_{\mu\nu}\,.
\end{equation}
Here, $T_{\mu\nu} = 2 \pd_{[\mu}T_{\nu]}$
is the exterior derivative of the clock one-form
and $\mathcal{K}_{\mu\nu} = - \frac{1}{2} \LL_V \Pi_{\mu\nu}$ is the extrinsic curvature.
The remaining terms in~\eqref{eq:pnr-lc-decomposition} are tensorial and are given by
\begin{subequations}
  \label{eq:pnr-gal-shift-tensors}
  \begin{align}
    \osb{C}{-2}^\rho{}_{\mu\nu}
    &= - T_{(\mu} \Pi^{\rho\sigma} T_{\nu)\sigma}\,,
    \label{eq:pnr-gal-shift-tensors-min2}
    \\
    \osb{C}{0}^\rho{}_{\mu\nu}
    &= V^\rho V^\sigma T_{(\mu} T_{\nu)\sigma}
    + \frac{1}{2} V^\rho T_{\mu\nu}\,,
    \label{eq:pnr-gal-shift-tensors-zero}
    \\
    \osb{C}{2}^\rho{}_{\mu\nu}
    &= - V^\rho \mathcal{K}_{\mu\nu}\,.
    \label{eq:pnr-gal-shift-tensors-plus2}
  \end{align}
\end{subequations}
Total covariant derivatives using the $\nablaCc_\rho$ connection satisfy
\begin{equation}
  \label{eq:pnr-connection-tot-der}
  \nablaCc_\mu X^\mu
  = \frac{1}{E} \pd_\mu \left(E\, X^\mu\right)
  - V^\mu T_{\mu\nu} X^\nu\,,
\end{equation}
which we will use frequently when integrating by parts.

\paragraph{Pre-non-relativistic Einstein--Hilbert decomposition.}
Using the decomposition of the Levi-Civita connection,
its Ricci scalar can then be decomposed in terms of the curvature of the Galilean connection~\eqref{eq:pnr-connection} together with other tensorial terms,
\begin{equation}
  \label{eq:pnr-ricci-scalar-on-shell-rewrite}
  \sqrt{-g}\, g^{\mu\nu} \osb{R}{LC}_{\mu\nu}
  \approx
  c\,E \left[
    \frac{c^2}{4} \Pi^{\mu\nu} \Pi^{\rho\sigma} T_{\mu\rho} T_{\nu\sigma}
    + \Pi^{\mu\nu} \osb{R}{\check{C}}_{\mu\nu}
    + \frac{1}{c^2} \left(\mK^{\mu\nu}\mK_{\mu\nu} - \mK^2\right)
  \right],
\end{equation}
which holds up to total exterior derivatives, corresponding to boundary terms in the action.
From this, we obtain the PNR rewriting of the Einstein--Hilbert action~\cite{Hansen:2020pqs},
\begin{equation}
  \label{eq:pnr-einstein-hilbert}
  \begin{split}
    S_\text{PNR}[T,\Pi]
    = \frac{c^4}{2\kappa} \int d^dx
    E \left[
      \frac{c^2}{4} \Pi^{\mu\nu} \Pi^{\rho\sigma} T_{\mu\rho} T_{\nu\sigma}
      + \Pi^{\mu\nu} \osb{R}{\check{C}}_{\mu\nu}
      + \frac{1}{c^2} \left(\mK^{\mu\nu}\mK_{\mu\nu} - \mK^2\right)
    \right]\,.
  \end{split}
\end{equation}
As we will discuss in Section~\ref{sec:exp-gal-lo-nlo}, this action is now rewritten in a form where each term can be Taylor expanded to obtain a Galilean action plus subleading corrections.

\paragraph{Pre-non-relativistic Einstein--Palatini decomposition.}
In the above, we rewrote the Einstein--Hilbert action in terms of the Galilean-adapted connection~\eqref{eq:pnr-connection} that was constructed from the decomposition~\eqref{eq:pnr-lc-decomposition} of the Levi-Civita connection.
Motivated by this decomposition, to obtain the corresponding reparametrization of the Einstein--Palatini action~\eqref{eq:einstein-palatini-action}, we similarly rewrite its general connection variable $\Gamma^\rho_{\mu\nu}$ as
\begin{align}
  \label{eq:pnr-lc-decomposition-off-shell}
  \Gamma^\rho_{\mu\nu}
  &= c^2\, \osb{C}{-2}^\rho{}_{\mu\nu}
  + C^\rho_{\mu\nu} + \osb{C}{0}^\rho{}_{\mu\nu}
  + \frac{1}{c^2} \osb{C}{2}^\rho_{\mu\nu}\,,
\end{align}
Here, the first and the two last terms are still the known tensors in~\eqref{eq:pnr-gal-shift-tensors}.
On the other hand, $C^\rho_{\mu\nu}$ is now an arbitrary connection variable.
As we will see in Section~\ref{ssec:pal-action-pnr-solve-conn}, it will be fixed to be the Galilean connection $\check{C}^\rho_{\mu\nu}$ in~\eqref{eq:pnr-connection} by its own equation of motion.

Under the change of variables~\eqref{eq:pnr-lc-decomposition-off-shell}, the Ricci tensor transforms following the general rule~\eqref{eq:general-riemann-tensor-conn-shift}, which gives
\begin{equation}
  \label{eq:pnr-ricci-decomposition-powers}
  R_{\mu\nu}
  = c^4\, \osb{R}{-4}_{\mu\nu}
  + c^2\, \osb{R}{-2}_{\mu\nu}
  + \osb{R}{0}_{\mu\nu}
  + \frac{1}{c^2} \osb{R}{2}_{\mu\nu}
  + \frac{1}{c^4} \osb{R}{4}_{\mu\nu}\,,
\end{equation}
where we grouped terms by their powers of $c$,
\begin{subequations}
  \label{eq:pnr-ricci-decomposition-powers-individual}
  \begin{align}
    \osb{R}{-4}_{\mu\nu}
    &= - \osb{C}{-2}^\rho{}_{\mu\lambda} \osb{C}{-2}^\lambda{}_{\rho\nu}
    + \osb{C}{-2}^\rho{}_{\rho\lambda} \osb{C}{-2}^\lambda{}_{\mu\nu}\,,
    \\
    \osb{R}{-2}_{\mu\nu}
    &= - \nablaC_\mu \osb{C}{-2}^\rho{}_{\rho\nu}
    + \nablaC_\rho \osb{C}{-2}^\rho{}_{\mu\nu}
    - \osb{T}{C}^\lambda{}_{\mu\rho} \osb{C}{-2}^\rho{}_{\lambda\nu}
    \\
    &{}\qquad\nonumber
    - \osb{C}{-2}^\rho{}_{\mu\lambda} \osb{C}{0}^\lambda{}_{\rho\nu}
    + \osb{C}{-2}^\rho{}_{\rho\lambda} \osb{C}{0}^\lambda{}_{\mu\nu}
    - \osb{C}{0}^\rho{}_{\mu\lambda} \osb{C}{-2}^\lambda{}_{\rho\nu}
    + \osb{C}{0}^\rho{}_{\rho\lambda} \osb{C}{-2}^\lambda{}_{\mu\nu}\,,
    \\
    \osb{R}{0}_{\mu\nu}
    &= \osb{R}{C}_{\mu\nu}
    - \nablaC_\mu \osb{C}{0}^\rho{}_{\rho\nu}
    + \nablaC_\rho \osb{C}{0}^\rho{}_{\mu\nu}
    - \osb{T}{C}^\lambda{}_{\mu\rho} \osb{C}{0}^\rho{}_{\lambda\nu}
    \\
    &{}\qquad\nonumber
    - \osb{C}{-2}^\rho{}_{\mu\lambda} \osb{C}{2}^\lambda{}_{\rho\nu}
    + \osb{C}{-2}^\rho{}_{\rho\lambda} \osb{C}{2}^\lambda{}_{\mu\nu}
    - \osb{C}{2}^\rho{}_{\mu\lambda} \osb{C}{-2}^\lambda{}_{\rho\nu}
    + \osb{C}{2}^\rho{}_{\rho\lambda} \osb{C}{-2}^\lambda{}_{\mu\nu}
    \\
    &{}\qquad\nonumber
    - \osb{C}{0}^\rho{}_{\mu\lambda} \osb{C}{0}^\lambda{}_{\rho\nu}
    + \osb{C}{0}^\rho{}_{\rho\lambda} \osb{C}{0}^\lambda{}_{\mu\nu}\,,
    \\
    \osb{R}{2}_{\mu\nu}
    &= - \nablaC_\mu \osb{C}{2}^\rho{}_{\rho\nu}
    + \nablaC_\rho \osb{C}{2}^\rho{}_{\mu\nu}
    - \osb{T}{C}^\lambda{}_{\mu\rho} \osb{C}{2}^\rho{}_{\lambda\nu}
    \\
    &{}\qquad\nonumber
    - \osb{C}{2}^\rho{}_{\mu\lambda} \osb{C}{0}^\lambda{}_{\rho\nu}
    + \osb{C}{2}^\rho{}_{\rho\lambda} \osb{C}{0}^\lambda{}_{\mu\nu}
    - \osb{C}{0}^\rho{}_{\mu\lambda} \osb{C}{2}^\lambda{}_{\rho\nu}
    + \osb{C}{0}^\rho{}_{\rho\lambda} \osb{C}{2}^\lambda{}_{\mu\nu}\,,
    \\
    \osb{R}{4}_{\mu\nu}
    &= - \osb{C}{2}^\rho{}_{\mu\lambda} \osb{C}{2}^\lambda{}_{\rho\nu}
    + \osb{C}{2}^\rho{}_{\rho\lambda} \osb{C}{2}^\lambda{}_{\mu\nu}\,.
  \end{align}
\end{subequations}
Using
the orthonormality properties~\eqref{eq:pnr-on-completeness}
as well as the fact that $V^\mu \mK_{\mu\nu}=0$,
and using the definitions~\eqref{eq:pnr-gal-shift-tensors} for the $\osb{C}{n}^\rho{}_{\mu\nu}$ terms that do not involve the connection,
we get
\begin{subequations}
  \label{eq:pnr-ricci-decomposition-powers-individual-simplified}
  \begin{align}
    \osb{R}{-4}_{\mu\nu}
    &= \frac{1}{4}
    T_\mu T_\nu \Pi^{\alpha\rho} \Pi^{\beta\sigma} T_{\alpha\beta} T_{\rho\sigma}\,,
    \\
    \osb{R}{-2}_{\mu\nu}
    &= \nablaC_\rho \osb{C}{-2}^\rho{}_{\mu\nu}
    - \osb{T}{C}^\lambda{}_{\mu\rho} \osb{C}{-2}^\rho{}_{\lambda\nu}
    + \frac{1}{2} T_\mu V^\rho \Pi^{\sigma\alpha} T_{\rho\sigma}  \left(
      T_{\alpha\nu}
      + T_\nu T_{\alpha\beta} V^\beta
    \right),
    \\
    \osb{R}{0}_{\mu\nu}
    &= \osb{R}{C}_{\mu\nu}
    - \nablaC_\mu \osb{C}{0}^\rho{}_{\rho\nu}
    + \nablaC_\rho \osb{C}{0}^\rho{}_{\mu\nu}
    - \osb{T}{C}^\lambda{}_{\mu\rho} \osb{C}{0}^\rho{}_{\lambda\nu}
    \\
    &{}\qquad\nonumber
    + \Pi^{\rho\sigma} K_{\rho(\mu} T_{\nu)\sigma}
    + T_{(\mu} K_{\nu)\rho} \Pi^{\rho\sigma} T_{\sigma\alpha} V^\alpha\,,
    \\
    \osb{R}{2}_{\mu\nu}
    &= \nablaC_\rho \osb{C}{2}^\rho{}_{\mu\nu}
    - \osb{T}{C}^\lambda{}_{\mu\rho} \osb{C}{2}^\rho{}_{\lambda\nu}\,,
    \\
    \osb{R}{4}_{\mu\nu}
    &= 0\,.
  \end{align}
\end{subequations}
We can simplify this further after contracting the Ricci tensor.
In particular, the Einstein--Palatini action~\eqref{eq:einstein-palatini-action} splits into powers of $c$ as follows,
\begin{align}
  S_\text{EP}
  &= \frac{c^3}{2\kappa} \int d^dx \sqrt{-g}\, g^{\mu\nu} R_{\mu\nu}
  = \frac{c^4}{2\kappa} \int d^dx E\, \left(
    - \frac{1}{c^2} V^\mu V^\nu + \Pi^{\mu\nu}
  \right) R_{\mu\nu}
  \\
  \label{eq:pnr-einstein-palatini-lagrangian-powers}
  &= \frac{1}{2\kappa} \int d^dx E\, \left[
    c^6\, \osb{\LL}{-6}
    + c^4\, \osb{\LL}{-4}
    + c^2\, \osb{\LL}{-2}
    + \osb{\LL}{0}
  \right].
\end{align}
Using the expressions in~\eqref{eq:pnr-ricci-decomposition-powers-individual-simplified} and now also working out the terms involving the covariant derivative and torsion, we then obtain
\begin{subequations}
  \label{eq:pnr-einstein-palatini-lagrangian-powers-individual}
  \begin{align}
    \label{eq:pnr-einstein-palatini-lagrangian-powers-individual-six}
    \osb{\LL}{-6}
    &= \frac{1}{4} \Pi^{\alpha\rho} \Pi^{\beta\sigma} T_{\alpha\beta} T_{\rho\sigma}\,,
    \\
    \label{eq:pnr-einstein-palatini-lagrangian-powers-individual-four}
    \osb{\LL}{-4}
    &= \Pi^{\mu\nu}
    \osb{R}{C}_{\mu\nu}
    + \osb{\nabla}{C}_\rho \left( \Pi^{\rho \nu} V^\sigma T_{\nu \sigma}\right)
    + \Pi^{\rho \nu} V^\sigma \osb{\nabla}{C}_\rho T_{\nu \sigma}
    \\
    \nonumber
    &{}\qquad
    + \Pi^{\rho \sigma} V^\mu \osb{T}{C}^\lambda{}_{\mu \rho} \left(
      \delta^{\nu}_\lambda - T_\lambda V^\nu
    \right)
    T_{\nu \sigma}
    + V^\mu V^\nu \Pi^{\rho \sigma} T_{\nu \sigma} \left(
      T_{\rho \mu} - \osb{T}{C}^{\alpha}{}_{\rho \mu} T_ \alpha
    \right),
    \\
    \osb{\LL}{-2}
    &= - V^\mu V^\nu \osb{R}{C}_{\mu\nu}
    - \Pi^{\mu\nu} \nablaC_\rho \left(V^\rho \mK_{\mu\nu}\right)
    + \Pi^{\mu\nu} \osb{T}{C}^\lambda{}_{\mu\rho} V^\rho \mK_{\lambda\nu}\,,
    \\
    \osb{\LL}{0}
    &= 0\,.
  \end{align}
\end{subequations}
As it turns out, the leading-order term in the rewriting of the Einstein--Palatini action is independent of the connection.
Furthermore, the final term
(corresponding to $V^\mu V^\nu \osb{R}{2}_{\mu\nu}$)
vanishes identically.
It would be interesting to understand these results from an algebraic perspective, classifying all possible curvature invariants including also torsion.
We will return to this question in the Discussion in Section~\ref{sec:discussion} below.
For now, we see that the pre-non-relativistic Einstein--Palatini action is
\begin{align}
  &S_\text{PNR-P}[T,\Pi,C]
  \nonumber
  \\
  \label{eq:pnr-palatini-action}
  &{}\qquad
  = \frac{c^4}{2\kappa} \int d^dx E\, \Bigg[
    \frac{c^2}{4} \Pi^{\alpha\rho} \Pi^{\beta\sigma} T_{\alpha\beta} T_{\rho\sigma}
    \\\nonumber
    &{}\qquad\qquad\qquad\qquad\quad
    + \Pi^{\mu\nu} \osb{R}{C}_{\mu\nu}
    + \osb{\nabla}{C}_\rho \left( \Pi^{\rho \nu} V^\sigma T_{\nu \sigma}\right)
    + \Pi^{\rho \nu} V^\sigma \osb{\nabla}{C}_\rho T_{\nu \sigma}
    \\\nonumber
    &{}\qquad\qquad\qquad\qquad\quad
    + V^\mu V^\nu \nablaC_\rho \left(\Pi^{\rho\sigma} T_\mu T_{\nu\sigma}\right)
    + \Pi^{\mu\nu} V^\rho V^\sigma T_{\nu\sigma} \nablaC_\rho T_\mu
    - \Pi^{\mu\nu} \nablaC_\mu \left(V^\rho T_{\rho\nu}\right)
    \\\nonumber
    &{}\qquad\qquad\qquad\qquad\quad
    + \frac{1}{c^2} \left(
      - V^\mu V^\nu \osb{R}{C}_{\mu\nu}
      - \Pi^{\mu\nu} \nablaC_\rho \left(V^\rho \mK_{\mu\nu}\right)
      + \Pi^{\mu\nu} \osb{T}{C}^\lambda{}_{\mu\rho} V^\rho \mK_{\lambda\nu}
    \right)
  \Bigg].
\end{align}
This action is equivalent to the Einstein--Palatini action~\eqref{eq:einstein-palatini-action-variation} and is written in terms of variables that are adapted to the non-relativistic $c\to\infty$ expansion with emerging Galilei structure.
Specifically, the action depends on the metric variables $T_\mu$ and $\Pi_{\mu\nu}$ introduced in~\eqref{eq:pnr-metric-decomposition},
as well as the arbitrary connection variable $C^\rho_{\mu\nu}$ introduced in~\eqref{eq:pnr-lc-decomposition-off-shell}.

\subsection{Solving for the Galilean connection}
\label{ssec:pal-action-pnr-solve-conn}
The form of the pre-non-relativistic (PNR) parametrization of the Einstein--Palatini action~\eqref{eq:pnr-palatini-action} differs significantly from the form of the PNR parametrization of the Einstein--Hilbert action~\eqref{eq:pnr-einstein-hilbert}.
In particular, in contrast to the Einstein--Palatini action~\eqref{eq:einstein-palatini-action}, it is not obtained simply by promoting the connection in the PNR Einstein--Hilbert action to a variable.
Nevertheless, we will now show that the PNR Einstein--Palatini action~\eqref{eq:pnr-palatini-action} reproduces the PNR Einstein--Hilbert action~\eqref{eq:pnr-einstein-hilbert} after solving for the connection.

One way to proceed is by directly varying the action with respect to the connection, as we did for the connection variable $\Gamma^\rho_{\mu\nu}$ in the original Einstein--Palatini action in Section~\ref{ssec:ep-review-solve-connection-direct}.
However, it is much easier to first perform a second field redefinition, similar to what we subsequently did in Section~\ref{ssec:ep-review-solve-connection-shift} above.
For this, we take
\begin{equation}
  \label{eq:pnr-second-shift}
  C^\rho_{\mu\nu}
  =\check{C}^\rho_{\mu\nu}
  + S^\rho{}_{\mu\nu}\,,
\end{equation}
which involves the Galilean-adapted connection $\check{C}^\rho_{\mu\nu}$ given in   \eqref{eq:pnr-connection}
and where $S^\rho{}_{\mu\nu}$ is an arbitrary tensor.
Combining this with the redefinition~\eqref{eq:pnr-lc-decomposition-off-shell}, we get
\begin{equation}
  \label{eq:pnr-lc-decomposition-and-second-shift}
  \Gamma^\rho_{\mu\nu}
  = \left(
    c^2\, \osb{C}{-2}^\rho{}_{\mu\nu}
    + \check{C}^\rho_{\mu\nu} + \osb{C}{0}^\rho{}_{\mu\nu}
    + \frac{1}{c^2} \osb{C}{2}^\rho_{\mu\nu}
  \right)
  = \osb{\Gamma}{LC}^\rho_{\mu\nu}
  + S^\rho{}_{\mu\nu}\,.
\end{equation}
In total, these two field redefinitions thus precisely reproduce the change of variables~\eqref{eq:connection-lc-shift} that we performed in Section~\ref{ssec:ep-review-solve-connection-shift}.
Consequently, after using~\eqref{eq:pnr-second-shift}, the PNR Palatini action~\eqref{eq:pnr-palatini-action} just becomes
\begin{align}
  \label{eq:pnr-einstein-palatini-double-shifted}
  S_\text{PNR-P}[T,\Pi,C]
  &\approx
  S_\text{PNR}[T,\Pi]
  \\\nonumber
  &{}\qquad
  +
  \frac{c^4}{2\kappa} \int d^dx E \left[
    - \frac{1}{c^2} V^\mu V^\nu + \Pi^{\mu\nu}
  \right] \left(
    - S^\rho{}_{\mu\lambda} S^\lambda{}_{\rho\nu}
    + S^\rho{}_{\rho\lambda} S^\lambda{}_{\mu\nu}
  \right).
\end{align}
This corresponds to the PNR decomposition of the Einstein--Palatini action in the form~\eqref{eq:einstein-palatini-shift-action}.
The combination in square brackets in the second term is nondegenerate, so we can solve the $S^\rho{}_{\mu\nu}$ equations of motion in exactly the same way as in Section~\ref{ssec:ep-review-solve-connection-shift}, raising and lowering indices just as we did with $g^{\mu\nu}$ there.
The equations of motion again lead to~\cite{Hansen:2020wqw}
\begin{equation}
  S^\rho{}_{\mu\nu}
  = A_\mu \delta^\rho_\nu\,,
\end{equation}
setting the shift tensor almost completely to zero, up to an ambiguity which drops out of the action~\eqref{eq:pnr-einstein-palatini-double-shifted}.
We can therefore solve for the connection in the PNR Einstein--Palatini action~\eqref{eq:pnr-palatini-action},
obtaining the Galilean connection $\check{C}^\rho_{\mu\nu}$ in~\eqref{eq:pnr-connection}
and reproducing the PNR Einstein--Hilbert action~\eqref{eq:einstein-palatini-shift-action} with the connection on shell.

So far, we have only focused on the connection variable and its equation of motion.
Indeed, while it is well-suited for that purpose, we do not want to use the action~\eqref{eq:pnr-einstein-palatini-double-shifted} for deriving the metric equations of motion, as this would just be equivalent to using the Einstein--Hilbert PNR action.
Instead, to obtain the metric equations of motion through a different path, we should take the variation directly in the PNR Palatini action~\eqref{eq:pnr-palatini-action}.
We will do so explicitly in Section~\ref{ssec:metric-eom-pal} below for the metric equations of motion of the next-to-leading order action in the Galilei expansion, which we construct in Section~\ref{sec:exp-gal-lo-nlo}.

\subsection{Pre-ultra-local Palatini action and Carroll connection}
\label{ssec:pal-action-pul}
Having obtained the Galilean pre-non-relativistic (PNR) form of the Palatini action in the previous sections, we now turn to the pre-ultra-local (PUL) parametrization which is adapted to the emerging Carroll symmetries.
For this, we use the same metric variables as defined in Equations~\eqref{eq:pnr-metric-decomposition} and~\eqref{eq:pnr-on-completeness}.
However, instead of the PNR decomposition of the Levi-Civita connection in~\eqref{eq:pnr-lc-decomposition}, we now use
\begin{align}
  \label{eq:pul-lc-decomposition}
  \osb{\Gamma}{LC}^\rho_{\mu\nu}
  &= \frac{1}{c^2} \osb{B}{-2}^\rho{}_{\mu\nu}
  + \tilde{C}^\rho_{\mu\nu} + \osb{B}{0}^\rho{}_{\mu\nu}
  + c^2\, \osb{B}{2}^\rho{}_{\mu\nu}\,.
\end{align}
Here, the $c^0$ terms are now split into a tensor $\osb{B}{0}^\rho{}_{\mu\nu}$ and
\begin{align}
  \label{eq:pul-connection}
  \tilde{C}^\rho_{\mu\nu}
  &= - V^\rho \pd_{(\mu} T_{\nu)} - V^\rho T_{(\mu} \LL_V T_{\nu)}
  \\
  &{}\qquad\nonumber
  + \frac{1}{2} \Pi^{\rho\lambda} \left[
    \pd_\mu \Pi_{\nu\lambda} + \pd_\nu \Pi_{\lambda\mu} - \pd_\lambda \Pi_{\mu\nu}
  \right]
  - \Pi^{\rho\lambda} T_\nu \mK_{\mu\lambda}.
\end{align}
which is a connection that is adapted to the Carrollian structure that emerges in the ultra-local $c\to0$ expansion.
Correspondingly, its covariant derivative
$\nablaCt_\rho$
satisfies
\begin{equation}
  \label{eq:pul-conn-met-comp-a}
  \nablaCt_\rho V^\mu = 0\,,
  \qquad
  \nablaCt_\rho \Pi_{\mu\nu} = 0\,.
\end{equation}
as well as
\begin{gather}
  \label{eq:pul-conn-met-comp-b}
  \nablaCt_\rho T_\mu
  = \frac{1}{2} T_{\mu\nu}
  - V^\rho T_{\rho(\mu} T_{\nu)}\,,
  \qquad
  \nablaCt_\rho \Pi^{\mu\nu}
  = - V^{(\mu} \Pi^{\nu)\sigma} T_{\sigma\lambda} \left[
    \delta^\lambda_\rho - V^\lambda T_\rho
  \right],
  \\
  \label{eq:pul-conn-torsion}
  \osb{T}{\tilde{C}}^\rho{}_{\mu\nu}
  = 2 \Pi^{\rho\lambda} T_{[\mu} \mK_{\nu]\lambda}\,.
\end{gather}
Finally, the tensors in~\eqref{eq:pul-lc-decomposition} are given by
\begin{subequations}
  \label{eq:pul-car-shift-tensors}
  \begin{align}
    \osb{B}{-2}^\rho{}_{\mu\nu}
    &= - V^\rho \mathcal{K}_{\mu\nu}\,,
    \label{eq:pul-car-shift-tensors-min2}
    \\
    \osb{B}{0}^\rho{}_{\mu\nu}
    &= \Pi^{\rho\lambda} T_\nu K_{\mu\lambda}\,,
    \label{eq:pul-car-shift-tensors-zero}
    \\
    \osb{B}{2}^\rho{}_{\mu\nu}
    &= - T_{(\mu} \Pi^{\rho\sigma} T_{\nu)\sigma}\,,
    \label{eq:pul-car-shift-tensors-plus2}
  \end{align}
\end{subequations}
Note that
$\osb{B}{-2}^\rho{}_{\mu\nu} = \osb{C}{2}^\rho{}_{\mu\nu}$
from~\eqref{eq:pnr-gal-shift-tensors-min2}
and
$\osb{B}{2}^\rho{}_{\mu\nu} = \osb{C}{-2}^\rho{}_{\mu\nu}$
from~\eqref{eq:pnr-gal-shift-tensors-plus2},
with the powers labeled differently because the expansions are opposite.
Finally, we have
\begin{equation}
  \label{eq:pul-connection-tot-der}
  \nablaCt_\mu X^\mu
  = \frac{1}{E} \pd_\mu \left(E\, X^\mu\right)
  - \mK T_\mu X^\mu\,,
\end{equation}
for total covariant derivatives using the $\nablaCt_\rho$ connection.
The resulting PUL reparametrization of the Einstein--Hilbert action is~\cite{Hansen:2021fxi}
\begin{equation}
  \label{eq:pul-einstein-hilbert}
  \begin{split}
    S_\text{PUL}[V,\Pi]
    \approx \frac{c^2}{2\kappa} \int d^dx
    E \left[
      \left(\mK^{\mu\nu}\mK_{\mu\nu} - \mK^2\right)
      + c^2 \Pi^{\mu\nu} \osb{R}{\tilde{C}}_{\mu\nu}
      + \frac{c^4}{4} \Pi^{\mu\nu} \Pi^{\rho\sigma} T_{\mu\rho} T_{\nu\sigma}
    \right]\,,
  \end{split}
\end{equation}
which again holds up to total derivative terms.

Following the PNR discussion around~\eqref{eq:pnr-lc-decomposition-off-shell}, we can now similarly decompose the general connection $\Gamma^\rho_{\mu\nu}$ entering in the Einstein--Palatini action~\eqref{eq:einstein-palatini-action} as follows,
\begin{align}
  \label{eq:pul-lc-decomposition-off-shell}
  \Gamma^\rho_{\mu\nu}
  &= \frac{1}{c^2} \osb{B}{-2}^\rho{}_{\mu\nu}
  + C^\rho_{\mu\nu} + \osb{B}{0}^\rho{}_{\mu\nu}
  + c^2\, \osb{B}{2}^\rho{}_{\mu\nu}\,,
\end{align}
where $C^\rho_{\mu\nu}$ is again an arbitrary connection.
Using~\eqref{eq:general-riemann-tensor-conn-shift}, this change of connection variables then leads to the following decomposition of the Ricci tensor,
\begin{align}
    R_{\mu\nu}
    =&  \frac{1}{c^2} \osb{R}{-2}_{\mu\nu}
    + \osb{R}{0}_{\mu\nu}
    + c^2\, \osb{R}{2}_{\mu\nu}
    + c^4\, \osb{R}{4}_{\mu\nu}\,,
    \label{eq:pul-ricci-decomposition-powers}
\end{align}
Consequently, the PUL Palatini action becomes
\begin{align}
    S_\text{PUL-P}[V,\Pi,C]
  =  \frac{1}{2\kappa} \int d^dx E\, \left[
    c^2\, \osb{\LL}{2}
    + c^4\, \osb{\LL}{4}
    + c^6\, \osb{\LL}{6}
  \right]. \label{eq:PUL-einstein-palatini-lagrangian-powers}
\end{align}
where each term is given by
\begin{align}
  \osb{\LL}{2}&= - V^\mu V^\nu \osb{R}{C}_{\mu \nu} - V^\mu V^\nu \osb{\nabla}{C}_{\rho} \left( \Pi^{\rho \lambda} T_\nu \mK_{\mu \lambda} \right) + V^{\mu} V^{\nu} \osb{\nabla}{C}_\mu \left(\mK T_{\nu} \right)  \nonumber
    \\
    &- \Pi^{\mu \nu} \osb{\nabla}{C}_{\rho} \left(V^{\rho} \mK_{\mu \nu} \right) + 2 \osb{T}{C}^\lambda{}_{\mu \rho} V^{\rho} \mK_{\lambda \nu} \Pi^{\mu \nu} +\mK^2  - \mK^{\mu \nu} \mK_{\mu \nu}\,,
    \\
    \osb{\LL}{4} &= \Pi^{\mu \nu} \osb{R}{C}_{\mu \nu} + \left( \Pi^{\mu \lambda} \Pi^{\nu \rho} \mK_{\rho \lambda} - \Pi^{\mu \nu} \mK \right) \osb{\nabla}{C}_{\mu} T_{\nu}  + V^\mu V^\nu \osb{\nabla}{C}_\rho \left( T_\mu \Pi^{\rho \sigma} T_{\nu \sigma} \right)  \nonumber
    \\
    &- V^\mu V^\nu \osb{T}{C}^{\lambda}{}_{\mu \rho} \Pi^{\rho \sigma} T_{(\lambda} T_{\nu)\sigma}\,,
    \\
    \osb{\LL}{6} &= \frac{1}{4} \Pi^{\alpha\rho} \Pi^{\beta\sigma} T_{\alpha\beta} T_{\rho\sigma} \,.
\end{align}
To see that this reduces to the PUL action~\eqref{eq:pul-einstein-hilbert}
after solving for the connection, we can follow Section~\ref{ssec:pal-action-pnr-solve-conn} and perform a second change of variables,
\begin{equation}
  \label{eq:pul-second-shift}
  C^\rho_{\mu\nu}
  = \tilde{C}^\rho_{\mu\nu} + S^\rho{}_{\mu\nu}\,,
\end{equation}
which involves the Carroll-adapted connection $\tilde{C}^\rho_{\mu\nu}$ given in   \eqref{eq:pul-connection}
and where $S^\rho{}_{\mu\nu}$ is again an arbitrary tensor.
Combining~\eqref{eq:pul-lc-decomposition-off-shell} and~\eqref{eq:pul-lc-decomposition}, the total change of variables with respect to the Einstein--Palatini action~\eqref{eq:einstein-palatini-action} is then
\begin{equation}
  \label{eq:pul-lc-decomposition-and-second-shift}
  \Gamma^\rho_{\mu\nu}
  = \frac{1}{c^2} \osb{B}{-2}^\rho{}_{\mu\nu}
    + \tilde{C}^\rho_{\mu\nu} + \osb{B}{0}^\rho{}_{\mu\nu}
    + c^2\, \osb{B}{2}^\rho{}_{\mu\nu}
  = \osb{\Gamma}{LC}^\rho_{\mu\nu}
  + S^\rho{}_{\mu\nu}\,,
\end{equation}
as in Equation~\eqref{eq:pnr-lc-decomposition-and-second-shift} for the PNR case.
As a result, the total reparametrization of the Einstein--Palatini action is
\begin{equation}
  \label{eq:pul-einstein-palatini-double-shifted}
  \begin{split}
    S_\text{EP}[g,\Gamma]
    &\approx
    S_\text{PUL}[V,\Pi]
    \\
    &{}\qquad
    +
    \frac{c^4}{2\kappa} \int d^dx E \left[
      - \frac{1}{c^2} V^\mu V^\nu + \Pi^{\mu\nu}
    \right] \left(
      - S^\rho{}_{\mu\lambda} S^\lambda{}_{\rho\nu}
      + S^\rho{}_{\rho\lambda} S^\lambda{}_{\mu\nu}
    \right).
  \end{split}
\end{equation}
The second line can again be solved to obtain $S^\rho{}_{\mu\nu}=A_\mu\delta^\rho_\nu$, which then drops out of the action.
Also in the PUL Carroll reparametrization, we thus see that we can obtain the PUL Einstein--Palatini action which reduces to the PUL Einstein--Hilbert action~\eqref{eq:pul-einstein-hilbert} after solving for the connection.

\section{Expanding the Palatini actions}
\label{sec:exp}
In the previous section, we have constructed the pre-non-relativistic (PNR) and pre-ultra-local (PUL) parametrizations of the Einstein--Palatini action.
Just like the PNR and PUL parametrizations of the Einstein--Hilbert action, they are adapted to the Galilean or Carrollian structures that emerge in the $c\to\infty$ and $c\to0$ expansion, respectively.
As we showed in Section~\ref{ssec:pal-action-pnr-solve-conn} for the PNR Palatini action and in Section~\ref{ssec:pal-action-pul} for the PUL Palatini action, their connection equations of motion lead to the desired result, and they reproduce the desired actions with the connection on shell.

We now focus on expanding the PNR and PUL Palatini actions, where both the metric and connection variables are expanded in powers of $1/c^2$ and $c^2$, respectively.
As we will see, the leading-order Galilean action resulting from the PNR action is independent of the connection.
The next-to-leading order does depend on the connection, and we solve the corresponding equation of motion to show that it reproduces the known next-to-leading order action for the metric variables.
In the PUL case, the leading-order Carroll action already depends on the connection, and we similarly show that it reduces to the correct result once the connection is put on shell.

\subsection{Leading-order and next-to-leading-order Galilei actions}
\label{sec:exp-gal-lo-nlo}
As a first step in the Galilean $c\to\infty$ expansion, we must expand the PNR metric variables that we introduced in~\eqref{eq:pnr-metric-decomposition}, which we parametrize as \cite{Hansen:2019pkl,Hansen:2020pqs}
\begin{align}
  T_\mu
  &= \tau_\mu + \frac{1}{c^2} m_\mu + \cdots\,,
  &
  V^\mu
  &= v^\mu + \cdots\,,
  \\
  \Pi_{\mu\nu}
  &= h_{\mu\nu} + \frac{1}{c^2} \Phi_{\mu\nu} + \cdots\,,
  &
  \Pi^{\mu\nu}
  &= h^{\mu\nu} + \cdots\, . 
\end{align}
The LO fields $(\tau_\mu,  h_{\mu\nu})$ together with the NLO fields $(m_\mu, \Phi_{\mu\nu})$ define what is
known as type II Newton-Cartan geometry.%
\footnote{Note that the fields $h_{\mu\nu}$ and $\Phi_{\mu\nu}$ transforms under local Galilean boosts, which arise from the expansion of local Lorentz boosts as mentioned at the start of Section~\ref{sec:pal-action}.
As discussed in for example~\cite{Hansen:2021fxi}, it can still be convenient to take this set of fields as the metric variables in the action in order to explicitly keep track of this boost symmetry.}
Further fields that appear beyond leading order are considered as additional gauge fields
living on this geometry. 
The vielbein determinant $E=e+\cdots$ is expanded accordingly, where we have
$e^2=-\det(-\tau_\mu\tau_\nu + h_{\mu\nu})$.
From the orthonormality relations~\eqref{eq:pnr-on-completeness} for the PNR variables,
we get
\begin{equation}
  \label{eq:on-completeness}
  v^\mu \tau_\mu = -1\,,
  \qquad
  v^\mu h_{\mu\nu} = 0\,,
  \qquad
  \tau_\mu h^{\mu\nu} = 0\,,
  \qquad
  \delta^\mu_\nu
  = - v^\mu \tau_\nu + h^{\mu\rho} h_{\rho\nu}\,.
\end{equation}
We have not introduced explicit notation for subleading variables in $V^\mu$ and $\Pi^{\mu\nu}$, since they can be solved in terms of $m_\mu$ and $\Phi_{\mu\nu}$ using the subleading terms obtained from the orthonormality relations~\eqref{eq:pnr-on-completeness}.
Additionally, we expand the connection variable,
\begin{equation}
  C^\rho_{\mu\nu}
  = \gamma^\rho_{\mu\nu}
  + \cdots\,.
\end{equation}
In the following, our goal will be to
solve for the leading-order connection $\gamma^\rho_{\mu\nu}$ and check that we recover the expansion of the PNR Einstein--Hilbert action~\eqref{eq:pnr-einstein-hilbert} when we plug the on-shell connection back into the action.
The expansion of the PNR Einstein--Hilbert action gives 
\begin{equation}
  S_\text{PNR}[T,\Pi]
  = c^6 S_\text{G-LO}[\tau,h]
  + c^4 S_\text{G-NLO}[\tau,h,m,\Phi] + \cdots\,.
\end{equation}
The leading-order and next-to-leading order actions in the $c\to\infty$ expansion are \cite{Hansen:2019pkl,Hansen:2020pqs}
\begin{align}
  \label{eq:gal-lo-action}
  S_\text{G-LO}[\tau,h]
  &= \frac{1}{2\kappa} \int d^dx\, e\,
  h^{\mu\rho} h^{\nu\sigma} \tau_{\mu\nu} \tau_{\rho\sigma}\,,
  \\
  \label{eq:gal-nlo-action}
  S_\text{G-NLO}[\tau,h,m,\Phi]
  &= \frac{1}{2\kappa} \int d^dx\, e\,
  \left[
    h^{\mu\nu} \check{R}_{\mu\nu}
    - 2 G^\mu_\tau m_\mu
    - G^{\mu\nu}_h \Phi_{\mu\nu}
  \right].
\end{align}
where $\tau_{\mu\nu} = 2\pd_{[\mu} \tau_{\nu]}$
and $\check{R}_{\mu\nu}$ is the Ricci tensor of the connection resulting from the leading-order term $\check\gamma^\rho_{\mu\nu}$ in the expansion of the PNR connection $\check{C}^\rho_{\mu\nu}$ in~\eqref{eq:pnr-connection},
\begin{equation}
  \label{eq:gal-connection}
  \check\gamma^\rho_{\mu\nu}
  = - v^\rho \pd_\mu \tau_\nu
  + \frac{1}{2} h^{\rho\sigma} \left(
    \pd_\mu h_{\nu\sigma} + \pd_\nu h_{\sigma\mu} - \pd_\sigma h_{\mu\nu}
  \right).
\end{equation}
As summarized in Appendix~\ref{sapp:geom-id-gal-conn},
this connection has the same metricity and torsion properties as $\check{C}^\rho_{\mu\nu}$, now formulated in terms of the leading-order metric variables $(\tau_\mu,h_{\mu \nu})$. 
Additionally,
$G^\mu_\tau$
and
$G^{\mu\nu}_h$
denote the $\tau_\mu$ and $h_{\mu\nu}$ equations of motion coming from the leading-order action~\eqref{eq:gal-lo-action}, respectively.
Consequently, we see that the equations of motion of the next-to-leading order fields $m_\mu$ and $\Phi_{\mu\nu}$ in the next-to-leading-order action~\eqref{eq:gal-nlo-action} precisely reproduce the equations of motion of their leading-order counterparts in the leading-order action~\eqref{eq:gal-lo-action}.
This pattern extends to arbitrary orders~\cite{Hansen:2019svu}.

We now want to consider the similar Galilean $c\to\infty$ expansion of the PNR Palatini action that we obtained in~\eqref{eq:pnr-palatini-action} above.
This leads to
\begin{equation}
  S_\text{PNR-P}[T,\Pi,C]
  = c^6 S_\text{GP-LO}[\tau,h,\gamma]
  + c^4 S_\text{GP-NLO}[\tau,h,m,\Phi,\gamma] + \cdots\,.
\end{equation}
At leading order, we obtain the action
\begin{align}
  \label{eq:gal-pal-lo-action}
  S_\text{GP-LO}[\tau,h]
  = \frac{1}{2\kappa} \int d^dx\, e\,
  h^{\mu\rho} h^{\nu\sigma} \tau_{\mu\nu} \tau_{\rho\sigma}\,,
\end{align}
which turns out to be independent of the connection and directly reproduces~\eqref{eq:gal-lo-action}.
After that, we get the following next-to-leading order (NLO) Galilean Palatini action,
\begin{align}
  \label{eq:gal-pal-nlo-action}
  S_\text{GP-NLO}[\tau,h,m,\Phi,\gamma]
  = \frac{1}{2\kappa}\int d^dx\, e\,
  &\Big[
    h^{\mu\nu} \osb{R}{\gamma}_{\mu\nu}
    - 2 G^\mu_\tau m_\mu
    - G^{\mu\nu}_h \Phi_{\mu\nu}
  \\\nonumber
  &{}\quad
    + \osb{\nabla}{\gamma}_\rho \left( h^{\rho \nu} v^\sigma \tau_{\nu \sigma}\right)
    + h^{\rho \sigma} v^\mu \osb{T}{\gamma}^\lambda_{\mu \rho} \left(
      \delta^{\nu}_\lambda - \tau_\lambda v^\nu
    \right)
  \\\nonumber
  &{}\quad
    + h^{\rho \nu} v^\sigma \osb{\nabla}{\gamma}_\rho \tau_{\nu \sigma}
    + v^\mu v^\nu h^{\rho \sigma} \tau_{\nu \sigma} \left(
      \tau_{\rho \mu} - \osb{T}{\gamma}^{\alpha}{}_{\rho \mu} \tau_ \alpha
    \right) 
  \Big]\,.
\end{align}
Here,
$\osb{R}{\gamma}_{\mu\nu}$
and
$\osb{T}{\gamma}^\rho{}_{\mu\nu}$
are the Ricci tensor and the torsion of
$\gamma^\rho_{\mu\nu}$,
the leading-order connection variable.
This action is not of the same form as the Galilean NLO action~\eqref{eq:gal-nlo-action},
but we will now show that the two agree once we have solved for the connection.

To obtain the connection equation of motion,
we can vary the action~\eqref{eq:gal-pal-nlo-action} directly with respect to the connection variable, as we did for the Einstein--Palatini action in Section~\ref{ssec:ep-review-solve-connection-direct}.
However, following our discussion for the Einstein--Palatini action in Section~\ref{ssec:ep-review-solve-connection-shift} and subsequently for the PNR Einstein--Palatini action in Section~\ref{ssec:pal-action-pnr-solve-conn},
it is easier to use the change of variables
\begin{equation}
  \label{eq:gal-pal-nlo-double-shift}
  \gamma^\rho_{\mu\nu}
  = \check\gamma^\rho_{\mu\nu}
  + s^\rho{}_{\mu\nu}\,,
\end{equation}
where $\check\gamma^\rho_{\mu\nu}$ is the leading-order Galilean connection~\eqref{eq:gal-connection}
and $s^\rho{}_{\mu\nu}$ is an arbitrary `shift' tensor.
After this substitution, the Galilean NLO Palatini action~\eqref{eq:gal-pal-nlo-action} becomes
\begin{equation}
  \label{eq:gal-pal-nlo-double-shifted}
  \begin{split}
    S_\text{GP-NLO}[\tau,h,m,\Phi,\gamma]
    &\approx
    S_\text{G-NLO}[\tau,h,m,\Phi]
    \\
    &{}\qquad
    + \frac{1}{2\kappa} \int d^dx e\, h^{\mu\nu}
    \left(
      - s^\rho{}_{\mu\lambda} s^\lambda{}_{\rho\nu}
      + s^\rho{}_{\rho\lambda} s^\lambda{}_{\mu\nu}
    \right),
  \end{split}
\end{equation}
which corresponds to the next-to-leading-order action~\eqref{eq:gal-nlo-action} plus a term involving the shift tensor.
The latter term only contains a contraction using the degenerate spatial metric~$h^{\mu\nu}$,
and we will therefore be able to solve for fewer components of the shift tensor than before.
However, all of the resulting ambiguities will still drop out of the action.
The equation of motion of $s^\rho{}_{\mu\nu}$ coming from~\eqref{eq:gal-pal-nlo-double-shifted} is
\begin{equation}
  \label{eq:gal-pal-nlo-double-shifted-eom}
  0
  =
  - h^{\mu\lambda} s^\nu{}_{\rho\lambda}
  - h^{\nu\lambda} s^\mu{}_{\lambda\rho}
  + \delta^\mu_\rho s^\nu{}_{\alpha\beta} h^{\alpha\beta}
  + h^{\mu\nu} s^\lambda{}_{\lambda\rho}\,.
\end{equation}
To solve this, we use the orthonormality relations~\eqref{eq:on-completeness} to note that the tensors
\begin{equation}
  h^\alpha_\mu = h^{\alpha\rho} h_{\rho\mu}\,,
  \qquad
  - v^\alpha \tau_\mu\,,
\end{equation}
square to themselves and sum to the identity.
As a result, they can be used to split the tangent bundle in `timelike' and `spacelike' components.
We denote timelike projections with a $0$ and spacelike projections with a dotted index, so that we have
\begin{equation}
  X_{\dot{\mu}} = h_\mu^\alpha X_\alpha\,,
  \qquad
  X_0 = v^\alpha X_\alpha\,,
  \qquad
  Y^{\dot{\mu}} = h^\mu_\alpha X^\alpha\,,
  \qquad
  Y^0 = \tau_\alpha X^\alpha\,.
\end{equation}
for a one-form $X_\mu$ and a vector $Y^\mu$.
In this way, we can decompose a given index in timelike and spacelike indices,
\begin{align}
  X_\mu
  = \delta_\mu^\alpha X_\alpha
  = - \tau_\mu X^0 + h_\mu^\alpha X_{\dot\alpha}\,,
  \qquad
  Y^\mu
  = \delta^\mu_\alpha Y^\alpha
  = - v^\mu Y_0 + h^\mu_\alpha Y^{\dot\alpha}\,.
\end{align}
Using these projections, the shift tensor $s^\rho{}_{\mu\nu}$ is split into eight components,
\begin{equation}
  \label{eq:lo-shift-tensor-decomposition}
  \begin{split}
    s^\rho{}_{\mu\nu}
    &= h^\rho_{\dot\gamma} h_\mu^{\dot\alpha} h_{\nu}^{\dot\beta}
    s^{\dot\gamma}{}_{\dot\alpha\dot\beta}
    - v^\rho h_\mu^{\dot\alpha} h_{\nu}^{\dot\beta}
    s^{0}{}_{\dot\alpha\dot\beta}
    - h^\rho_{\dot\gamma} \tau_\mu h_{\nu}^{\dot\beta}
    s^{\dot\gamma}{}_{0\dot\beta}
    - h^\rho_{\dot\gamma} h_\mu^{\dot\alpha} \tau_\nu
    s^{\dot\gamma}{}_{\dot\alpha0}
    \\
    &{}\qquad
    + v^\rho \tau_\mu h_{\nu}^{\dot\beta}
    s^{0}{}_{0\dot\beta}
    + v^\rho h_\mu^{\dot\alpha} \tau_\nu
    s^{0}{}_{\dot\alpha0}
    + h^\rho_{\dot\gamma} \tau_\mu \tau_\nu
    s^{\dot\gamma}{}_{00}
    - v^\rho \tau_\mu \tau_\nu
    s^{0}{}_{00}\,.
  \end{split}
\end{equation}
We can then solve for most of these components by taking the timelike and spacelike projections of the $\mu\nu$ indices of the equation of motion~\eqref{eq:gal-pal-nlo-double-shifted-eom}.
In total, this leads to
\begin{align}
  s^\rho{}_{\mu\nu}
  &= \delta^\rho_\nu A_\mu
  - h^\rho_\nu \tau_\mu s^{\dot\alpha}{}_{\dot\alpha0}
  + \tau_\mu s_{\dot\nu}{}^{\dot\rho}{}_0
  - \tau_\nu s^{\dot\rho}{}_{\dot\mu0}
  + \tau_\mu \tau_\nu s^{\dot\rho}{}_{00}\,.
  \label{eq:gal-pal-nlo-shift-tensor-answer}
\end{align}
The first term is just the ambiguity we already saw in Sections~\ref{sec:ep-review}
and~\ref{ssec:pal-action-pnr-solve-conn} above,
and the other terms are additional ambiguities.
However, all of them still drop out of the Galilean NLO Palatini action~\eqref{eq:gal-pal-nlo-double-shifted}.
After putting the connection on shell, we therefore recover precisely the Galilean NLO action~\eqref{eq:gal-nlo-action} as desired.

\subsection{Leading-order Carroll action}
\label{sec:exp-car-lo}
We now turn to the Carroll case, restricting for simplicity to the leading-order action. 
Similar to what we did for the Galilean case we must expand the PUL metric variables.
We write the leading-order terms in the $c \rightarrow 0$ expansion of the variables in~\eqref{eq:pnr-metric-decomposition} as
\begin{align}
  T_\mu
  &= \tau_\mu + \cdots\,,
  &
  V^\mu
  &= v^\mu + \cdots\,,
  \\
  \Pi_{\mu\nu}
  &= h_{\mu\nu}  + \cdots\,,
  &
  \Pi^{\mu\nu}
  &= h^{\mu\nu} + \cdots\, , 
\end{align}
where the dots now denote terms of ${\cal{O}} (c^2)$. 
Although the notation is similar, the leading-order fields now define a Carroll geometry and transform under local Carroll boosts, which descend from local Lorentz boosts.
Again, the vielbein determinant is expanded as $E=e+\cdots$,
where $e^2=-\det(-\tau_\mu \tau_\nu + h_{\mu\nu})$,
and the orthonormality relations~\eqref{eq:pnr-on-completeness} for the PUL variables lead to
\begin{equation}
  \label{eq:on-completeness-repeat}
  v^\mu \tau_\mu = -1\,,
  \qquad
  v^\mu h_{\mu\nu} = 0\,,
  \qquad
  \tau_\mu h^{\mu\nu} = 0\,,
  \qquad
  \delta^\mu_\nu
  = - v^\mu \tau_\nu + h^{\mu\rho} h_{\rho\nu}\,.
\end{equation}
Expanding the PUL Einstein--Hilbert action~\eqref{eq:pul-einstein-hilbert} to leading order gives
\begin{equation}
  S_\text{PUL}[V,\Pi]
  = c^2 S_\text{C-LO}[v,h]+\cdots\,,
\end{equation}
where the leading-order Carroll gravity action is~\cite{Hansen:2021fxi}
\begin{equation}
  S_\text{C-LO}
  = \frac{1}{2\kappa} \int d^dx e
  \left(K^{\mu\nu} K_{\mu\nu} - K^2\right),
  \label{eq:car-lo-action}
\end{equation}
and $K_{\mu \nu} = - \frac{1}{2} \mathcal{L}_v h_{\mu \nu}$ is the leading-order extrinsic curvature.
This action is also known as the `electric' Carroll gravity action, and it has appeared in various forms throughout the literature~\cite{Henneaux:1979vn,Hartong:2015xda,Henneaux:2021yzg}.

We now want to recover this action from the leading-order term in our PUL Einstein--Palatini action~\eqref{eq:PUL-einstein-palatini-lagrangian-powers}.
For this, we also expand the connection variable,
\begin{equation}
   C^\rho_{\mu\nu}
  = \gamma^\rho_{\mu\nu}
  + \cdots\,.
\end{equation}
so that we get the expansion
\begin{equation}
  S_\text{PUL-P}[V,\Pi,C]
  = c^2 S_\text{CP-LO}[v,h,\gamma] +\cdots\,,
\end{equation}
where the leading-order Carroll Palatini action is
\begin{align} 
  S_\text{CP-LO}[v,h,\gamma]
  &=\frac{1}{2\kappa} \int d^dx e\, \Big(
    - v^\mu v^\nu \osb{R}{\gamma}_{\mu \nu}
    - v^\mu v^\nu \osb{\nabla}{\gamma}_{\rho} \left[ h^{\rho \lambda} \tau_\nu K_{\mu \lambda} \right]
    \nonumber
    \\
    &{}\qquad\qquad\qquad\qquad
    + v^{\mu} v^{\nu} \osb{\nabla}{\gamma}_\mu \left[K \tau_{\nu} \right]
    - h^{\mu \nu} \osb{\nabla}{\gamma}_{\rho} \left[v^{\rho} K_{\mu \nu} \right]
  \label{eq:car-pal-lo-action}
    \\
    &{}\qquad\qquad\qquad\qquad
    + 2 \osb{T}{\gamma}^\lambda{}_{\mu \rho} v^{\rho} K_{\lambda \nu} h^{\mu \nu}
  - K^{\mu \nu} K_{\mu \nu} +K^2 \Big).
  \nonumber
\end{align}
To obtain its connection equation of motion, we again use a change of variables,
\begin{equation}
  \label{eq:car-pal-lo-double-shift}
   \gamma^\rho_{\mu\nu}
  = \tilde \gamma^\rho_{\mu\nu}
  + s^\rho{}_{\mu\nu}\,,
\end{equation}
where $s^\rho{}_{\mu\nu}$ is an arbitrary tensor
and $\tilde \gamma^\rho_{\mu\nu}$ is the leading-order Carroll connection
from the expansion of the PUL connection~$\tilde{C}^\rho_{\mu\nu}$ in~\eqref{eq:pul-connection},
\begin{align}
  \tilde \gamma^\rho_{\mu\nu} 
  &= - v^\rho \pd_{(\mu} \tau_{\nu)} - v^\rho \tau_{(\mu} \LL_v \tau_{\nu)}
  \\
  &{}\qquad\nonumber
  + \frac{1}{2} h^{\rho\lambda} \left[
    \pd_\mu h_{\nu\lambda} + \pd_\nu h_{\lambda\mu} - \pd_\lambda h_{\mu\nu}
  \right]
  - h^{\rho\lambda} \tau_\nu K_{\mu\lambda}\,. 
\end{align}
This connection inherits the torsion and Carroll metric-compatibility properties,
\begin{gather}
  \label{eq:car-conn-met-comp-a}
  \tilde\nabla_\rho v^\mu
  = 0\,,
  \qquad
  \tilde\nabla_\rho h_{\mu\nu}
  = 0\,,
  \\
  \label{eq:car-conn-met-comp-b}
  \tilde\nabla_\rho \tau_\mu
  = \frac{1}{2} \tau_{\mu\nu}
  - v^\rho \tau_{\rho(\mu} \tau_{\nu)}\,,
  \qquad
  \tilde\nabla_\rho h^{\mu\nu}
  = - v^{(\mu} h^{\nu)\sigma} \tau_{\sigma\lambda} \left[
    \delta^\lambda_\rho - v^\lambda \tau_\rho
  \right],
  \\
  \label{eq:car-conn-torsion}
  \tilde{T}^\rho{}_{\mu\nu}
  = 2 h^{\rho\lambda} \tau_{[\mu} K_{\nu]\lambda}\,.
\end{gather}
After the change of variables~\eqref{eq:car-pal-lo-double-shift}, the Carroll LO Palatini action~\eqref{eq:car-pal-lo-action} becomes
\begin{equation}
  \label{eq:car-pal-lo-double-shifted}
  \begin{split}
    S_\text{CP-LO}[v,h,\gamma]
    &\approx S_\text{C-LO}[v,h]
    + \frac{1}{2\kappa} \int d^dx e\, v^\mu v^\nu
    \left(
      - s^\rho{}_{\mu\lambda} s^\lambda{}_{\rho\nu}
      + s^\rho{}_{\rho\lambda} s^\lambda{}_{\mu\nu}
    \right),
  \end{split}
\end{equation}
which corresponds to the leading-order term in the expansion of the PUL Einstein--Hilbert action in the form~\eqref{eq:pul-einstein-palatini-double-shifted}.
The resulting equation of motion for $s^\rho{}_{\mu \nu}$ is now
\begin{equation}
  \label{eq:car-pal-lo-double-shifted-eom}
  0
  =
  - v^\mu v^\lambda s^\nu{}_{\rho\lambda}
  - v^\nu v^\lambda s^\mu{}_{\lambda\rho}
  + \delta^\mu_\rho s^\nu{}_{\alpha\beta} v^\alpha v^\beta
  + v^\mu v^\nu s^\lambda{}_{\lambda\rho}\,.
\end{equation}
which we can solve in the language introduced around~\eqref{eq:lo-shift-tensor-decomposition} to find
\begin{align}
  s^\rho{}_{\mu\nu}
  &= \delta^{\rho}_\nu A_\mu + v^\rho \tau_\mu s^0{}_{0 \dot{\nu}} - v^\rho s^0{}_{\dot{\mu} \dot{\nu}} + F^{\dot{\rho}}{}_{\dot{\mu} \dot{\nu}} - \frac{1}{3} h^{\rho}_\mu F^{\dot{\lambda}}{}_{\dot{\lambda} \dot{\nu}}\,,
  \label{eq:car-pal-lo-shift-tensor-answer}
\end{align}
where $F^{\dot{\lambda}}{}_{\dot{\mu} \dot{\nu}}$ is an arbitrary spatial tensor that is traceless over its first and last index.
Again, we recover the ambiguity $A_\mu$ that we already saw in Sections~\ref{sec:ep-review} and~\ref{sec:exp-gal-lo-nlo}.
Furthermore, due to the degenerate $v^\mu v^\nu$ tensor entering in the action
we get additional ambiguities,
which nevertheless still drop out of the action.
Remarkably,
even though the leading-order Carroll Palatini action~\eqref{eq:car-pal-lo-action} depends quite nontrivially on the connection,
we still recover precisely the connection-independent leading-order `electric' Carroll gravity action~\eqref{eq:car-lo-action} after putting the connection on shell.
This result could perhaps help evade the claimed `no-go' theorem for constructing the electric Carroll gravity action from a gauging procedure~\cite{Figueroa-OFarrill:2022mcy}, see also the Discussion in Section~\ref{sec:discussion} below.

\section{Metric equations of motion from NLO Galilei action}
\label{sec:metric-eom}
In the previous sections, we showed how to systematically obtain a Palatini-type action for both the Galilean and Carroll expansions of general relativity.
We also checked that the resulting actions reduce to the known expressions once the connection variables are on shell.
Up to now, however, we have not yet considered the metric equations of motion.

As we briefly reviewed in Section~\ref{sec:ep-review}, the Einstein--Palatini action has the remarkable feature that it is of exactly the same form as the Einstein--Hilbert action, the only difference being that the connection is promoted to a variable in the former.
The derivation of the metric equation of motion is almost exactly the same in both actions, as the variation of the Ricci tensor only contributes a boundary term.
However, as we saw above, the form of the Palatini actions for the Galilean and Carrollian expansion of GR are significantly different from the actions that they reduce to once the connection is on shell.
As a result, while they still reduce to the correct answer with the connection on shell, the form of the metric equations of motion coming from the Palatini action will now be significantly different.

We illustrate this with a concrete example from the Galilean expansion.
As we saw in Section~\ref{sec:exp-gal-lo-nlo}, the leading-order Galilean Palatini action turns out to be independent of the connection.
We therefore consider the next-to-leading order Palatini action~\eqref{eq:gal-pal-nlo-action},
\begin{align}
  \label{eq:gal-pal-nlo-action-repeat}
  S_\text{GP-NLO}[\tau,h,m,\Phi,\gamma]
  = \frac{1}{2\kappa}\int d^dx\, e\,
  &\Big[
    h^{\mu\nu} \osb{R}{\gamma}_{\mu\nu}
    - 2 G^\mu_\tau m_\mu
    - G^{\mu\nu}_h \Phi_{\mu\nu}
  \\\nonumber
  &{}\quad
    + \osb{\nabla}{\gamma}_\rho \left( h^{\rho \nu} v^\sigma \tau_{\nu \sigma}\right)
    + h^{\rho \sigma} v^\mu \osb{T}{\gamma}^\lambda_{\mu \rho} \left(
      \delta^{\nu}_\lambda - \tau_\lambda v^\nu
    \right)
  \\\nonumber
  &{}\quad
    + h^{\rho \nu} v^\sigma \osb{\nabla}{\gamma}_\rho \tau_{\nu \sigma}
    + v^\mu v^\nu h^{\rho \sigma} \tau_{\nu \sigma} \left(
      \tau_{\rho \mu} - \osb{T}{\gamma}^{\alpha}{}_{\rho \mu} \tau_ \alpha
    \right) 
  \Big]\,.
\end{align}
As we saw previously, using this action we can solve for the connection~$\gamma^\rho_{\mu\nu}$, and plugging the answer back in reproduces the Galilean NLO action~\eqref{eq:gal-nlo-action},
\begin{equation}
  \label{eq:gal-nlo-action-repeat}
  S_\text{G-NLO}[\tau,h,m,\Phi]
  = \frac{1}{2\kappa} \int d^dx\, e
  \left[
    h^{\mu\nu} \check{R}_{\mu\nu}
    - 2 G^\mu_\tau m_\mu
    - G^{\mu\nu}_h \Phi_{\mu\nu}
  \right].
\end{equation}
We now consider the variation of both actions with respect to the metric data.
First, in Section~\ref{ssec:metric-eom-so} we review the known variation of~\eqref{eq:gal-nlo-action-repeat}, where the Ricci tensor depends on the metric data through the fixed Galilean connection~$\check\gamma^\rho_{\mu\nu}$ given in
\eqref{eq:gal-connection}.
In contrast, when varying the action~\eqref{eq:gal-pal-nlo-action-repeat}, the Ricci tensor is independent of the metric data and only depends on the connection variable.
The same holds for the torsion tensors and covariant derivatives entering in the action.
The resulting metric equations of motion are derived in Section~\ref{ssec:metric-eom-pal}, and we show that they reduce to the equations derived directly from the second-order action once the connection is put on shell.
This demonstrates that our Palatini approach provides an alternative path to the metric equations of motion, which is especially useful for the spatial $h_{\mu\nu}$ equations of motion.

\subsection{From the second-order action}
\label{ssec:metric-eom-so}
Let us first review the metric equations of motion that were obtained from varying the second-order action~\eqref{eq:gal-nlo-action-repeat} in~\cite{Hansen:2020pqs}.
First, note that $G^\mu_\tau$ and $G^{\mu\nu}_h$, which correspond to the $\tau_\mu$ and $h_{\mu\nu}$ equations of motion of the leading-order action~\eqref{eq:gal-lo-action}, do not depend on the connection.
Their explicit form is given in~\cite{Hansen:2020pqs}, but they turn out to be equivalent to
\begin{equation}
  \label{eq:ttnc}
  h^{\mu\rho} h^{\nu\sigma} \tau_{\rho\sigma} = 0\,,
\end{equation}
which is also known as the twistless torsional Newton--Cartan (TTNC) condition.
Geometrically, this implies that $\tau_\mu$ can be used to define equal-time surfaces.
Following the general pattern of nested actions and equations of motion mentioned in Section~\ref{sec:exp-gal-lo-nlo}, we see that the $m_\mu$ and $\Phi_{\mu\nu}$ equations of motion in the next-to-leading-order action~\eqref{eq:gal-nlo-action-repeat} are equivalent to the same condition~\eqref{eq:ttnc}.

Next, we consider the $\tau_\mu$ and $h_{\mu\nu}$ equations of motion.
They contain several different types of contributions,
\begin{align}
  2\kappa\, \delta \LL_\text{G-NLO}
  &= \delta e
  \left[
    h^{\mu\nu} \check{R}_{\mu\nu}
    - 2 G^\mu_\tau m_\mu
    - G^{\mu\nu}_h \Phi_{\mu\nu}
  \right]
  - e \left(
    2 \delta G^\mu_\tau m_\mu
    + \delta G^{\mu\nu}_h \Phi_{\mu\nu}
  \right)
  + e\, \delta h^{\mu\nu} \check{R}_{\mu\nu}
  \nonumber
  \\
  \label{eq:gal-nlo-metric-eom-so-var}
  &{}\qquad
  + e\, h^{\mu\nu} \delta\check{R}_{\mu\nu}\,.
\end{align}
The terms on the first line of this expression lead to the same result whether the connection is allowed to vary or not.
Therefore, our interest is mainly in the term on the last line, as this will not be present in the metric equations of motion if the connection is an independent variable.
In the present case, we can use the variation of the connection~$\check{\gamma}^\rho_{\mu\nu}$
from Equation~\eqref{eq:gal-connection-variation-split} together with the identity
\begin{equation}
  \label{eq:check-ricci-variation-contraction}
  h^{\mu\nu} \delta \check{R}_{\mu\nu}
  = - \check\nabla_\mu \left( h^{\mu\nu} \delta \check\gamma^\rho_{\rho\nu} \right)
  + \check\nabla_\rho \left( h^{\mu\nu} \delta \check\gamma^\rho_{\mu\nu} \right)
  + h^{\mu\nu} v^\lambda \tau_{\mu\rho} \delta \check\gamma^\rho_{\lambda\nu}\,.
\end{equation}
Note that the first two terms are not simply boundary terms due to the identity~\eqref{eq:gal-connection-tot-der-app} for a total covariant $\check\nabla_\mu$ derivative.
In total, we obtain
\begin{subequations}
  \begin{align}
    \label{eq:check-ricci-variation-contraction-time-answer}
    h^{\mu\nu} \delta_\tau\check{R}_{\mu\nu}
    &\approx
    v^\rho \tau_{\rho\alpha} \left(
      K^{\alpha\beta} - K h^{\alpha\beta}
    \right) \delta \tau_\beta\,,
    \\
    \label{eq:check-ricci-variation-contraction-space-answer}
    h^{\mu\nu} \delta_h\check{R}_{\mu\nu}
    &\approx
    \left(
      h^{\mu\rho} h^{\nu\sigma} - h^{\mu\nu} h^{\rho\sigma}
    \right)
    v^\alpha
    \left(
      v^\beta \tau_{\alpha\mu} \tau_{\beta\nu}
      + \check\nabla_\nu \tau_{\alpha\mu}
    \right) \delta h_{\rho\sigma}\,,
  \end{align}
\end{subequations}
where we have used the TTNC condition~\eqref{eq:ttnc} from the $m_\mu$ and $\Phi_{\mu\nu}$ equations of motion to simplify the expressions.
In the following, our goal will be to obtain these variations from the additional terms that appear in the Palatini NLO Galilean action~\eqref{eq:gal-pal-nlo-action-repeat}.

\subsection{From the Palatini action}
\label{ssec:metric-eom-pal}
We can write the Palatini next-to-leading order Galilean action~\eqref{eq:gal-pal-nlo-action-repeat} as
\begin{gather}
  \label{eq:gal-pal-nlo-action-repeat-compact}
  S_\text{GP-NLO}[\tau,h,m,\Phi,\gamma]
  = \frac{1}{2\kappa}\int d^dx\, e\,
  \left[
    h^{\mu\nu} \osb{R}{\gamma}_{\mu\nu}
    - 2 G^\mu_\tau m_\mu
    - G^{\mu\nu}_h \Phi_{\mu\nu}
    + \Sigma
  \right],
\end{gather}
where the additional terms compared to the form of the Galilean NLO action~\eqref{eq:gal-nlo-action-repeat} are
\begin{equation}
  \begin{split}
    \Sigma
    &=
    \osb{\nabla}{\gamma}_\rho \left( h^{\rho \nu} v^\sigma \tau_{\nu \sigma}\right)
    + h^{\rho \sigma} v^\mu \osb{T}{\gamma}^\lambda_{\mu \rho} \left(
      \delta^{\nu}_\lambda - \tau_\lambda v^\nu
    \right)
    \\
    &{}\qquad
    + h^{\rho \nu} v^\sigma \osb{\nabla}{\gamma}_\rho \tau_{\nu \sigma}
    + v^\mu v^\nu h^{\rho \sigma} \tau_{\nu \sigma} \left(
      \tau_{\rho \mu} - \osb{T}{\gamma}^{\alpha}{}_{\rho \mu} \tau_ \alpha
    \right).
  \end{split}
\end{equation}
Again, varying with respect to the next-to-leading-order metric data $m_\mu$ and $\Phi_{\mu\nu}$
just returns the leading-order equations of motion $G^\mu_\tau=0$ and $G^{\mu\nu}_h=0$
which are equivalent to the TTNC condition~\eqref{eq:ttnc}.
Varying the leading-order metric data $\tau_\mu$ and $h_{\mu\nu}$ gives
\begin{align}
  2\kappa\, \delta \LL_\text{GP-NLO}
  &= \delta e
  \left[
    h^{\mu\nu} \osb{R}{\gamma}_{\mu\nu}
    - 2 G^\mu_\tau m_\mu
    - G^{\mu\nu}_h \Phi_{\mu\nu}
  \right]
  - e \left(
    2 \delta G^\mu_\tau m_\mu
    + \delta G^{\mu\nu}_h \Phi_{\mu\nu}
  \right)
  + e\, \delta h^{\mu\nu} \osb{R}{\gamma}_{\mu\nu}
  \nonumber
  \\
  \label{eq:gal-nlo-metric-eom-pal-var}
  &{}\qquad
  +  \delta e\, \Sigma
  + e\, \delta \Sigma\,.
\end{align}
Upon putting the connection on shell, we see that the first line of this variation
reproduces the first line of the variation~\eqref{eq:gal-nlo-metric-eom-so-var} in the second-order action.

Our goal now is therefore to check that the second line of~\eqref{eq:gal-nlo-metric-eom-pal-var}
is equivalent to the variations~\eqref{eq:check-ricci-variation-contraction-time-answer} and
\eqref{eq:check-ricci-variation-contraction-space-answer} of the Ricci tensor from $\delta \LL_\text{G-NLO}$.
Using~\eqref{eq:nc-measure-variation}, the first term gives
\begin{equation}
  \label{eq:gal-nlo-metric-eom-pal-var-de-sigma}
  \delta e\, \Sigma
  = e \left(
    2 v^\rho \delta \tau_\rho - h^{\rho\sigma} \delta h_{\rho\sigma}
  \right)
  h^{\mu\nu}
  v^\alpha
  \left(
    v^\beta \tau_{\alpha\mu} \tau_{\beta\nu}
    + \check\nabla_\nu \tau_{\alpha\mu}
  \right).
\end{equation}
Here and in the following, we have taken the connection to be on shell after performing the variation.
Then let us focus on the $e\,\delta_\tau \Sigma$ variation.
Note that we can write
\begin{equation}
  \delta_\tau \tau_{\mu\nu}
  = 2 \osb{\nabla}{\gamma}_{[\mu} \delta\tau_{\nu]}
  + \osb{T}{\gamma}^\rho{}_{\mu\nu} \delta\tau_\lambda\,.
\end{equation}
With this, the $\tau_\mu$ variations of the individual terms in $\Sigma$ give
\begin{align}
  \delta_\tau \left[
    \osb{\nabla}{\gamma}_\rho \left( h^{\rho \nu} v^\sigma \tau_{\nu \sigma}\right)
  \right] 
  &\approx
  - K h^{\mu\rho} v^\nu \tau_{\mu\nu} \delta \tau_\rho
  - v^\mu v^\nu h^{\rho\sigma} \check\nabla_\mu \tau_{\nu\sigma} \delta\tau_\rho
  \\
  &{}\nonumber\qquad
  - v^\rho v^\nu h^{\mu\sigma} \tau_{\nu\sigma} \check\nabla_\mu \delta\tau_\rho\,,
  \\
  \delta_\tau \left[ h^{\rho \nu} v^\sigma \osb{\nabla}{\gamma}_\rho \tau_{\nu \sigma} \right]
  &\approx
  - v^\mu K^{\rho\sigma} \tau_{\mu\sigma} \delta\tau_\rho
  - v^\mu v^\nu h^{\rho\sigma} \check\nabla_\mu \tau_{\nu\sigma} \delta\tau_\rho
  \\
  &{}\nonumber\qquad
  + v^\mu v^\nu h^{\rho\sigma} \tau_{\mu\sigma} \check\nabla_\nu \delta\tau_\rho\,,
  \\
  \delta_\tau \left[ v^\mu v^\nu h^{\rho \sigma} \tau_{\nu \sigma} \left(
      \tau_{\rho \mu} - \osb{T}{\gamma}^{\alpha}{}_{\rho \mu} \tau_ \alpha
  \right) \right]
  &=
  - \left(
    v^\mu v^\nu h^{\rho\sigma} 
    + h^{\mu\nu} v^\rho v^\sigma
  \right)
  \tau_{\nu\sigma} \check\nabla_\mu \delta\tau_\rho\,,
  \\
  \delta_\tau \left[ h^{\rho \sigma} v^\mu \osb{T}{\gamma}^\lambda_{\mu \rho} \left(
      \delta^{\nu}_\lambda - \tau_\lambda v^\nu
    \right)
  \tau_{\nu \sigma} \right]
  &=
  - 2 \left(
    v^\mu v^\nu h^{\rho\sigma} 
    + h^{\mu\nu} v^\rho v^\sigma
  \right)
  \tau_{\nu\sigma} \check\nabla_\mu \delta\tau_\rho\,,
\end{align}
We have integrated by parts using~\eqref{eq:gal-connection-tot-der-app} in the first two terms.
Together, this sums to
\begin{align}
  e\, \delta_\tau\Sigma
  &\approx
  - 2e\, v^\rho h^{\mu\nu} v^\alpha
  \left(
    v^\beta \tau_{\alpha\mu} \tau_{\beta\nu}
    + \check\nabla_\nu \tau_{\alpha\mu}
  \right) \delta\tau_\rho
  + e \, v^\rho \tau_{\rho\alpha} \left(
    K^{\alpha\beta} - K h^{\alpha\beta}
  \right) \delta \tau_\beta\,,
\end{align}
after additional integration by parts.
Together with the contribution from~\eqref{eq:gal-nlo-metric-eom-pal-var-de-sigma}
this precisely reproduces the remaining contributions~\eqref{eq:check-ricci-variation-contraction-time-answer} from the $\tau_\mu$ equation of motion above.
Next, the individual terms in the $e\,\delta_h\Sigma$ variation give
\begin{align}
  e\, \delta_h \left[
    \osb{\nabla}{\gamma}_\rho \left(
      h^{\rho \nu} v^\sigma \tau_{\nu \sigma}
    \right)
  \right] 
  &\approx -e\, v^\mu v^\nu h^{\rho \alpha} h^{\sigma \beta}
  \tau_{\mu \rho} \tau_{\nu \sigma} \delta h_{\alpha \beta}\,,
  \\
  e\, \delta_h \left[ h^{\rho \nu} v^\sigma \osb{\nabla}{\gamma}_\rho \tau_{\nu \sigma} \right]
  &= - e\, h^{\rho \alpha} h^{\sigma \beta} v^\nu  \check \nabla_\rho \tau_{\sigma \nu}  \delta h_{\alpha \beta},
  \\
  e\, \delta_h \left[
    v^\mu v^\nu h^{\rho \sigma} \tau_{\nu \sigma} \left(
      \tau_{\rho \mu} - \osb{T}{\gamma}^{\alpha}{}_{\rho \mu} \tau_ \alpha
    \right)
  \right]
  &=
  0\,,
  \\
  e\, \delta_h \left[ h^{\rho \sigma} v^\mu \osb{T}{\gamma}^\lambda_{\mu \rho} \left(
      \delta^{\nu}_\lambda - \tau_\lambda v^\nu
    \right)
  \tau_{\nu \sigma} \right]
  &=
  2e\, v^\mu v^\nu
  h^{\rho \alpha} h^{\sigma \beta} \tau_{\mu \rho} \tau_{\nu \sigma} \delta h_{\alpha \beta}\,,
\end{align}
where we have used the TTNC condition~\eqref{eq:ttnc}.
These variations add to
\begin{equation}
  e\,\delta_h \Sigma
  =
  h^{\mu\rho} h^{\nu\sigma}
  v^\alpha
  \left(
    v^\beta \tau_{\alpha\mu} \tau_{\beta\nu}
    + \check\nabla_\nu \tau_{\alpha\mu}
  \right) \delta h_{\rho\sigma}\,.
\end{equation}
Together with the contribution from~\eqref{eq:gal-nlo-metric-eom-pal-var-de-sigma}, we see that we reproduce the desired result~\eqref{eq:check-ricci-variation-contraction-space-answer} to get the correct $h_{\mu\nu}$ equations of motion.

We have therefore verified that the metric equations of motion for the next-to-leading-order Galilean Palatini action~\eqref{eq:gal-pal-nlo-action-repeat-compact} indeed reproduce the equations of motion from the corresponding second-order action~\eqref{eq:gal-nlo-action-repeat} once the connection is put on shell.

\section{Discussion and outlook}
\label{sec:discussion}

We conclude the paper with a brief discussion of some open problems.

First of all, it would be interesting and important to further compare our approach and results with related 
`first-order' methods in the literature \cite{Bergshoeff:2017btm,Campoleoni:2022ebj,Figueroa-OFarrill:2022mcy,Bergshoeff:2023rkk}, including the application of the gauging procedure. 
In particular, it seems the gauging procedure misses terms in the action that correspond to torsion. 
For both the Galilean and Carrollian case these are the LO terms in the expansion, which we have seen are captured by our method.
This may thus point at a way in which to circumvent the no-go result for obtaining the leading-order or `electric' Carrollian theory from a gauging procedure~\cite{Figueroa-OFarrill:2022mcy}, and similarly for the leading TTNC term of the Galilean case. 
Additionally, it would be interesting to check if the ambiguities we found in solving for the Galilean and Carrollian connection variables not only drop out of the action but also do not affect the corresponding point particle geodesics, as was argued in~\cite{Dadhich:2012htv,Bernal:2016lhq} for the ambiguity resulting from the Einstein--Palatini action.

Subsequently, we would like to be able to use such a gauging procedure to systematically obtain subleading actions in the non-relativistic and ultra-local expansions of general relativity.
Relatedly, a natural next step would be to use the approach in this paper to obtain the NNLO term for the Galilean Palatini action and the NLO term for the Carroll Palatini action respectively.
For the Galilean NLO theory, we saw a simplification in the Palatini approach to computing the metric equations of motion.
However, for the Carroll LO theory, the computation would actually be more involved than the variation of the original action.
While obtaining and understanding the equations of motion of the full Carroll NLO theory (completing the `magnetic' truncation already considered in~\cite{Hansen:2021fxi}) remains an important open problem, it is therefore uncertain if our Palatini approach is truly helpful for this.

In addition, there are a number of direct applications of the Palatini formalism developed here. 
For the NLO Galilean case, obtained in this paper, it would be interesting to use the Palatini form to compute the charge (mass)
of the strong-coupling Schwarzschild solution \cite{VandenBleeken:2017rij,Hansen:2019vqf,Adam:2009gq} which is a solution 
of non-relativistic gravity with non-zero torsion.
For the Carroll case, an extra motivation to determine the NLO equations of motion 
is to further elucidate the appearance of mass in the Carroll black hole solution.
These solutions were observed~\cite{Hansen:2021fxi,Perez:2022jpr} in a particular truncation of the NLO theory, known as the magnetic Carroll theory, but understanding them in the full NLO theory would be useful.
Additionally, it would be interesting to see how they are connected to the black hole-like solutions with angular and linear momentum (but no mass) that were observed in the LO theory~\cite{Hansen:2021fxi}.
Another direction connects to the fact that the Schwarzschild geometry close to the singularity is described by a LO Carroll solution that is of Kasner type \cite{Oling:2021,deBoer:2023fnj}.
One could therefore hope to apply the NLO Carroll theory and beyond to shed further light on the near-singularity dynamics of general relativity, providing subleading corrections in the Belinski-Khalatnikov-Lifshitz limit \cite{Belinski:2017fas}.

It would also be nice to explore the connection to the more general `geometrical trinity' description of general relativity, using torsion, non-metricity or curvature (see for example~\cite{BeltranJimenez:2019esp}) as well as its non-relativistic version~\cite{Wolf:2023rad}.
Finally, we emphasize that the expansions considered in this paper are done in terms of even powers of the speed of light, which is a consistent subsector of GR.
It could be interesting to consider a generalization to odd powers following~\cite{Ergen:2020yop}.

\subsection*{Acknowledgements}
We are thankful to Francesco Alessio and Benjamin Knorr for useful discussions, and to Jelle Hartong for useful discussions and helpful comments on an early version of this paper.
JM thanks Nordita and NO thanks Edinburgh University for hospitality. 
The work of NO and GO is supported in part by VR project grant 2021-04013. The work of NO is also supported by the Villum Foundation Experiment project 00050317, ``Exploring the wonderland of Carrollian physics: Extreme gravity, spacetime horizons and supersonic fluids''. Nordita is supported in part by Nordforsk. 
The work of GO is also supported by the Royal Society URF of Jelle Hartong through the Research Fellows Enhanced Research Expenses 2022 (RF\textbackslash ERE\textbackslash 221013).

\appendix
\section{Conventions and useful geometric identities}
\label{app:geom-id}
Below we list some definitions and we give some identities that will be useful in several parts of the main text.
Unless otherwise specified, the following holds for arbitrary connections.
Our conventions follow those of~\cite{Hansen:2020pqs,Hansen:2021fxi}.

\subsection{General connection and curvature}
\label{sapp:geom-id-gen-conn}
For an arbitrary covariant derivative $\nabla_\mu$, we define its curvature using
\begin{subequations}
  \label{eq:general-riemann-tensor-def}
  \begin{align}
    \label{eq:general-riemann-tensor-def-from-vector}
    [\nabla_{\mu},\nabla_{\nu}] X^\rho
    &= - R_{\mu\nu\sigma}{}^\rho X^\sigma
    - T^\sigma{}_{\mu\nu} \nabla_\sigma X^\rho\,,
    \\
    \label{eq:general-riemann-tensor-def-from-covector}
    [\nabla_{\mu},\nabla_{\nu}] Y_\sigma
    &= R_{\mu\nu\sigma}{}^\rho Y_\rho
    - T^\rho{}_{\mu\nu} \nabla_\rho Y_\sigma\,.
  \end{align}
\end{subequations}
In terms of the connection components $\Gamma^\rho_{\mu\nu}$,
the Riemann curvature tensor and the torsion tensor are given by
\begin{align}
  \label{eq:general-riemann-tensor-def-cpts}
  R_{\mu\nu\sigma}{}^\rho
  &= - \pd_\mu \Gamma_{\nu\sigma}^\rho + \pd_\nu \Gamma_{\mu\sigma}^\rho
  - \Gamma^\rho_{\mu\lambda} \Gamma^\lambda_{\nu\sigma}
  + \Gamma^\rho_{\nu\lambda} \Gamma^\lambda_{\mu\sigma}\,,
  \\
  \label{eq:general-torsion-tensor-def-cpts}
  T^\rho{}_{\mu\nu}
  &= 2 \Gamma^\rho_{[\mu\nu]}\,.
\end{align}
These definitions hold for any connection, independent of its metricity or torsion properties.
In particular, they do not assume the existence of a Riemannian metric.

\paragraph{Connection shift.}
We often want to transition from a given connection
$\Gamma^\rho_{\mu\nu}$
to another connection
$\Gamma'^\rho_{\mu\nu}$
using relations of the form
\begin{equation}
  \label{eq:general-conn-shift}
  \Gamma^\rho_{\mu\nu}
  = \Gamma'^\rho_{\mu\nu} + S^\rho{}_{\mu\nu}\,.
\end{equation}
Here, the `shift' $S^\rho{}_{\mu\nu}$ between the two connections is a tensor.
The Riemann tensor, the torsion tensor and the covariant derivative of a tensor $X^\mu{}_\nu$ transform as
\begin{align}
  \label{eq:general-riemann-tensor-conn-shift}
  R_{\mu\nu\sigma}{}^\rho
  &= R'_{\mu\nu\sigma}{}^\rho
  - \nabla'_\mu S^\rho{}_{\nu\sigma}
  + \nabla'_\nu S^\rho{}_{\mu\sigma}
  - T'^\lambda{}_{\mu\nu} S^\rho{}_{\lambda\sigma}
  - S^\rho{}_{\mu\lambda} S^\lambda{}_{\nu\sigma}
  + S^\rho{}_{\nu\lambda} S^\lambda{}_{\mu\sigma}\,,
  \\
  \label{eq:general-torsion-tensor-conn-shift}
  T^\rho{}_{\mu\nu}
  &= T'^\rho{}_{\mu\nu}
  + 2 S^\rho{}_{[\mu\nu]}\,,
  \\
  \label{eq:general-cov-der-conn-shift}
  \nabla_\rho X^\mu{}_\nu
  &= \nabla'_\rho X^\mu{}_\nu
  + S^\mu{}_{\rho\sigma} X^\sigma{}_\nu
  - S^\sigma{}_{\rho\nu} X^\mu{}_\sigma\,,
\end{align}
where $R'_{\mu\nu\sigma}{}^\rho$, $\nabla'_\mu$ and $T'^\rho{}_{\mu\nu}$
are the Riemann curvature, covariant derivative and torsion tensor of the
$\Gamma'^\rho_{\mu\nu}$ connection.
Of course, the transformation of the covariant derivative of $X^\mu{}_\nu$ can be extended to arbitrary tensors by linearity.

\paragraph{Connection variation.}
We also often consider variations in the connection,
\begin{equation}
  \label{eq:general-conn-va}
  \Gamma^\rho_{\mu\nu}
  \to \Gamma^\rho_{\mu\nu} + \delta\Gamma^\rho_{\mu\nu}\,.
\end{equation}
The resulting variations of the Riemann tensor, torsion tensor and covariant derivative are then
\begin{align}
  \label{eq:general-riemann-tensor-conn-var}
  \delta R_{\mu\nu\sigma}{}^\rho
  &= - \nabla_\mu \delta\Gamma^\rho_{\nu\sigma}
  + \nabla_\nu \delta\Gamma^\rho_{\mu\sigma}
  - T^\lambda{}_{\mu\nu} \delta\Gamma^\rho_{\lambda\sigma}\,,
  \\
  \label{eq:general-torsion-tensor-conn-var}
  \delta T^\rho{}_{\mu\nu}
  &= 2 \delta\Gamma^\rho_{[\mu\nu]}\,,
  \\
  \label{eq:general-cov-der-conn-var}
  \delta \left(\nabla_\rho X^\mu{}_\nu\right)
  &= \delta\Gamma^\mu_{\rho\sigma} X^\sigma{}_\nu
  - \delta\Gamma^\sigma_{\rho\nu} X^\mu{}_\sigma\,,
\end{align}
which corresponds to the linear terms of the expressions for a general shift in (\ref{eq:general-riemann-tensor-conn-shift})-(\ref{eq:general-cov-der-conn-shift}).

\paragraph{Derivative of tensor densities.}
For a tensor density $z^\mu{}_\nu$ of weight $w$, we have
\begin{equation}
  \label{eq:tensor-density-cov-der}
  \nabla_\rho z^\mu{}_\nu
  = \pd_\rho z^\mu{}_\nu
  + \Gamma^\mu_{\rho\sigma} z^\sigma{}_\nu
  - \Gamma^\sigma_{\rho\nu} z^\mu{}_\sigma
  - w\, \Gamma^\sigma_{\rho\sigma} z^\mu{}_\nu,
\end{equation}
which again extends linearly to tensors with other index structures.
In particular, for a vector density~$j^\mu$ of weight $+1$ we have
\begin{equation}
  \label{eq:weight-one-vector-density-divergence}
  \nabla_\mu j^\mu
  = \pd_\mu j^\mu
  + \Gamma^\mu_{\mu\alpha} j^\alpha
  - \Gamma^\alpha_{\mu\alpha} j^\mu
  = \pd_\mu j^\mu + T^\alpha{}_{\alpha\mu} j^\mu,
\end{equation}
which we will use in several places in the main text.

\subsection{Properties of the Galilean and Carrollian metric variables}
\label{sapp:geom-id-gal-car-met-var}
As we briefly review at the beginning of Section~\ref{sec:pal-action},
we obtain Galilean and Carrollian metric variables from the leading-order terms in the corresponding expansion of the Lorentzian metric.
In both cases, the resulting metric data can be described using
$(\tau_\mu, h_{\mu\nu})$
and their inverse
$(v^\mu, h^{\mu\nu})$,
which satisfy the orthonormality relations
\begin{equation}
  \label{eq:on-completeness-app}
  v^\mu \tau_\mu = -1\,,
  \qquad
  v^\mu h_{\mu\nu} = 0\,,
  \qquad
  \tau_\mu h^{\mu\nu} = 0\,,
  \qquad
  \delta^\mu_\nu
  = - v^\mu \tau_\nu + h^{\mu\rho} h_{\rho\nu}\,.
\end{equation}
The corresponding integration measure is given by
$e^2 = \det(\tau_\mu\tau_\nu + h_{\mu\nu})$.
These metric variables will transform under local Galilean or Carrollian boost transformations (which follow from the corresponding limits of the local Lorentz boosts), but we will not need the details of this for purposes of this paper.

Instead, let us list the relations between the variations of the metric data,
\begin{subequations}
  \label{eq:nc-variations-relations}
  \begin{gather}
    \delta v^\mu
    = v^\mu v^\rho \delta \tau_\rho
    - h^{\mu\rho} v^\sigma \delta h_{\rho\sigma}\,,
    \qquad
    \delta h^{\mu\nu}
    = 2 v^{(\mu} h^{\nu)\rho} \delta \tau_\rho
    - h^{\mu\rho} h^{\nu\sigma} \delta h_{\rho\sigma}\,,
    \\
    \delta \tau_\mu
    = \tau_\mu \tau_\rho \delta v^\rho
    - h_{\mu\rho} \tau_\sigma \delta h^{\rho\sigma}\,,
    \qquad
    \delta h_{\mu\nu}
    = 2 \tau_{(\mu} h_{\nu)\rho} \delta v^\rho
    - h_{\mu\rho} h_{\nu\sigma} \delta h^{\rho\sigma}\,,
  \end{gather}
\end{subequations}
which can be obtained directly from~\eqref{eq:on-completeness-app}.
Additionally, we have
\begin{equation}
  \label{eq:nc-measure-variation}
  \delta e
  = e \left(- v^\mu \delta \tau_\mu + \frac{1}{2} h^{\mu\nu} \delta h_{\mu\nu} \right)
  = e \left(\tau_\mu \delta v^\mu - \frac{1}{2} h_{\mu\nu} \delta h^{\mu\nu}\right).
\end{equation}
for the variation of the integration measure.
We will mostly vary our expressions with respect to the variables
$(\tau_\mu, h_{\mu\nu})$,
and the above expressions~\eqref{eq:nc-variations-relations} then allow us to express the corresponding variations of their inverse metric objects in terms of $(\delta \tau_\mu, \delta h_{\mu\nu})$.

\subsection{Properties of the Galilean connection}
\label{sapp:geom-id-gal-conn}
We collect some properties of the Galilean connection~$\check\gamma^\rho_{\mu\nu}$
introduced in Equation~\eqref{eq:gal-connection},
\begin{equation}
  \label{eq:gal-connection-app}
  \check\gamma^\rho_{\mu\nu}
  = - v^\rho \pd_\mu \tau_\nu
  + \frac{1}{2} h^{\rho\sigma} \left(
    \pd_\mu h_{\nu\sigma} + \pd_\nu h_{\sigma\mu} - \pd_\sigma h_{\mu\nu}
  \right).
\end{equation}
Its metricity properties and its torsion tensor are given by
\begin{gather}
  \label{eq:gal-connection-metricity-app}
  \check\nabla_\rho \tau_\mu
  = 0 \,,
  \quad
  \check\nabla_\rho h^{\mu\nu}
  = 0 \,,
  \quad
  \check\nabla_\rho v^\mu
  = - h^{\mu\nu} \mK_{\nu\rho}\,,
  \quad
  \check\nabla_\rho h_{\mu\nu}
  = -2 \tau_{(\mu} K_{\nu)\rho}\,,
  \\
  \check{T}^\rho{}_{\mu\nu}
  = - v^\rho \tau_{\mu\nu}\,.
\end{gather}
Here, $K_{\mu\nu} = - \frac{1}{2} \LL_v h_{\mu\nu}$ is the extrinsic curvature
and $\tau_{\mu\nu} = 2 \pd_{[\mu} \tau_{\nu]}$.
The corresponding total exterior derivative satisfies
\begin{equation}
  \label{eq:gal-connection-tot-der-app}
  \check\nabla_\mu X^\mu
  = \frac{1}{e} \pd_\mu \left(e\, X^\mu\right)
  - v^\mu \tau_{\mu\nu} X^\nu\,.
\end{equation}
All this is equivalent to the leading-order terms in the expansion of
the properties of the pre-non-relativistic connection $\check{C}^\rho_{\mu\nu}$ listed in Section~\ref{ssec:pal-action-pnr}.
Finally, we get
\begin{subequations}
  \label{eq:gal-connection-variation-split}
  \begin{align}
    \label{eq:gal-connection-variation-split-tau}
    \delta_\tau \check\gamma^\rho_{\mu\nu}
    &= - v^\rho \check\nabla_\mu \delta\tau_\nu
    + K_{\mu\nu} h^{\rho\lambda} \delta\tau_\lambda\,,
    \\
    \label{eq:gal-connection-variation-split-hdown}
    \delta_h \check\gamma^\rho_{\mu\nu}
    &= \frac{h^{\rho\lambda}}{2} \left[
      \check\nabla_\mu \delta h_{\nu\lambda}
      + \check\nabla_\nu \delta h_{\lambda\mu}
      - \check\nabla_\lambda \delta h_{\mu\nu}
    \right]
    \\
    &{}\qquad\nonumber
    + h^{\rho\alpha} \tau_{\alpha(\mu} v^\beta \delta h_{\nu)\beta}
    + \frac{1}{2} h^{\rho\alpha} \tau_{\mu\nu} v^\beta \delta h_{\alpha\beta}\,,
  \end{align}
\end{subequations}
upon varying the connection with respect to the metric variables.

\section{Solving the connection for the Einstein--Palatini action}
\label{app:solving-connection-for-ep}
In this appendix, we give some computational details for our brief review of the Einstein--Palatini action in Section~\ref{sec:ep-review}.
In particular, following Section~\ref{ssec:pal-action-pnr}, we first solve the connection equation of motion that is obtained by varying the connection directly,
following the discussion in \S21.2 of~\cite{Misner:1973prb} (which we extend here to non-zero torsion).
We then solve the equation of motion for the shift tensor introduced in Section~\ref{ssec:ep-review-solve-connection-shift}, which provides an easier path to the same answer, following for example~\cite{Dadhich:2012htv}.
As always, we assume that our connection variables are completely general to begin with, so in particular we do not require vanishing torsion or metricity.

\paragraph{Solving the EOM obtained by varying directly.}
In Section~\ref{ssec:ep-review-solve-connection-direct}, we varied the Einstein--Palatini action with respect to the connection variable $\Gamma^\rho_{\mu\nu}$ and obtained the equation of motion~\eqref{eq:einstein-palatini-direct-gamma-eom},
\begin{equation}
  \label{eq:einstein-palatini-direct-gamma-eom-app}
  \begin{split}
    0
    &=
    \delta^\mu_\rho \nabla_\alpha \left(\sqrt{-g}\, g^{\alpha\nu} \right)
    - \nabla_\rho \left(\sqrt{-g}\, g^{\mu\nu}\right)
    \\
    &{}\qquad
    - \sqrt{-g}\, T^\alpha{}_{\alpha\beta} g^{\beta\nu} \delta_\rho^\mu
    + \sqrt{-g}\, T^\alpha{}_{\alpha\rho} g^{\mu\nu}
    - \sqrt{-g}\, T^\mu{}_{\alpha\rho} g^{\alpha\nu}\,.
  \end{split}
\end{equation}
Roughly speaking, we can interpret this as an equation for the torsion
$T^\rho{}_{\mu\nu} = 2 \Gamma^\rho_{[\mu\nu]}$
and the non-metricity, which is defined as
\begin{equation}
  \label{eq:down-non-metricity}
  Q_{\rho\mu\nu} = \nabla_\rho g_{\mu\nu}\,.
\end{equation}
As is well known, these two tensors together fully determine a given connection $\Gamma^\rho_{\mu\nu}$,
\begin{align}
  \Gamma^\rho_{\mu\nu}
  &= \osb{\Gamma}{LC}^\rho_{\mu\nu}
  + K^\rho{}_{\mu\nu} + L^\rho{}_{\mu\nu}\,,
  \label{eq:general-connection-K-L}
\end{align}
where the first term is the usual Levi-Civita connection, and the two other terms are known as the contorsion and disformation tensors,
\begin{gather}
  \osb{\Gamma}{LC}^\rho_{\mu\nu}
  = \frac{g^{\rho\sigma}}{2} \left[
    \pd_\mu g_{\nu\sigma} + \pd_\nu g_{\sigma\mu} - \pd_\sigma g_{\mu\nu}
  \right],
  \\
  \label{eq:contortion-disformation}
  K^\rho{}_{\mu\nu}
  = - \frac{1}{2} \left[
    T_{\mu\nu}{}^\rho - T_\nu{}^\rho{}_\mu - T^\rho{}_{\mu\nu}
  \right],
  \qquad
  L^\rho{}_{\mu\nu}
  = - \frac{1}{2} \left[
    Q_{\mu\nu}{}^\rho + Q_\nu{}^\rho{}_\mu - Q^\rho{}_{\mu\nu}
  \right].
\end{gather}
Note that we are lowering and raising indices using the Lorentzian metric and its inverse.
To be precise, the equation of motion~\eqref{eq:einstein-palatini-direct-gamma-eom-app} is an equation for the torsion tensor and the inverse `metricity density'
\begin{equation}
  q_\rho{}^{\mu\nu}
  = \nabla_\rho \left(E\, g^{\mu\nu}\right).
\end{equation}
It is convenient to introduce the following notation for their traces,
\begin{gather}
  \label{eq:app-direct-traces-def}
  a_\mu
  = E\, T_\mu{}^\rho{}_\rho
  = 0\,,
  \qquad
  b_\mu
  = E\, T_{\rho\mu}{}^\rho\,,
  \qquad
  c_\mu
  = E\, T_\rho{}^\rho{}_\mu
  = - b_\mu\,,
  \\
  \mathcal{A}_\mu
  = q_\mu{}^\rho{}_\rho\,,
  \qquad
  \mathcal{B}_\mu
  = q_{\rho\mu}{}^\rho\,,
  \qquad
  \mathcal{C}_\mu
  = q_\rho{}^\rho{}_\mu
  = \mathcal{B}_\mu\,.
\end{gather}
Note that these are all one-form densities of weight one.
Using this notation and raising the $\rho$ index, the equation of motion~\eqref{eq:einstein-palatini-direct-gamma-eom-app} becomes
\begin{equation}
  \label{eq:einstein-palatini-direct-gamma-eom-app-symbolic}
  0
  = g^{\mu\rho} \mathcal{B}^\nu - q^{\rho\mu\nu}
  + b^\nu g^{\mu\rho} - b^\rho g^{\mu\nu}
  - E\, T^{\mu\nu\rho}\,.
\end{equation}
We first want to solve for as many of the traces as possible.
By contracting the equation over $\mu\nu$ and $\mu\rho$ we get
\begin{equation}
  \mathcal{B}_\mu = - \frac{(d-2)}{(d-1)} b_\mu\,,
  \qquad
  \mathcal{A}_\mu = - \frac{d(d-2)}{(d-1)} b_\mu\,.
\end{equation}
Then~\eqref{eq:einstein-palatini-direct-gamma-eom-app-symbolic} gives the metricity density in terms of the torsion and its traces,
\begin{equation}
  \label{eq:einstein-palatini-direct-gamma-eom-app-symbolic-for-q}
  q^{\rho\mu\nu}
  = - g^{\mu\rho} \frac{(d-2)}{(d-1)} b^\nu
  + b^\nu g^{\mu\rho}
  - b^\rho g^{\mu\nu}
  - E T^{\mu\nu\rho}\,.
\end{equation}
Antisymmetrizing this over $\mu\nu$ and subtracting and adding the cyclic permutations,
this expression allows us to solve for the torsion tensor
\begin{equation}
  \label{eq:palatini-direct-gamma-eom-torsion-result}
  T^\rho{}_{\mu\nu}
  = \frac{E\inv}{(d-1)} \left[
    \delta^\rho_\nu b_\mu
    - \delta^\rho_\mu b_\nu
  \right]
  = \delta^\rho_\nu A_\mu - \delta^\rho_\mu A_\nu\,,
  \qquad
  A_\mu
  = \frac{E\inv}{(d-1)} b_\mu\,.
\end{equation}
Plugging this back into~\eqref{eq:einstein-palatini-direct-gamma-eom-app-symbolic-for-q}, we obtain the non-metricity density
\begin{equation}
  \label{eq:palatini-direct-gamma-eom-metricity-density-result}
  q_\rho{}^{\mu\nu}
  = (2-d) E\, A_\rho g^{\mu\nu}\,,
\end{equation}
or equivalently the (inverse) non-metricity tensor
$\nabla_\rho g^{\mu\nu}= 2 A_\rho g^{\mu\nu}$.
Following the general identity~\eqref{eq:general-connection-K-L} determining a connection in terms of its non-metricity and torsion tensors, this leads to
\begin{equation}
  \label{eq:einstein-palatini-direct-gamma-sol-app}
  \Gamma^\rho_{\mu\nu}
  = \osb{\Gamma}{LC}^\rho_{\mu\nu} + \delta^\rho_\nu A_\mu\,,
\end{equation}
as claimed in Equation~\eqref{eq:einstein-palatini-direct-gamma-sol} in the main text.

\paragraph{Solving the EOM obtained from a shift.}
In Section~\ref{ssec:ep-review-solve-connection-shift} we first used a change of variables
from the arbitrary connection variable $\Gamma^\rho_{\mu\nu}$
to the Levi-Civita connection and an arbitrary `shift' tensor $S^\rho{}_{\mu\nu}$,
\begin{equation}
  \label{eq:connection-lc-shift-app}
  \Gamma^\rho_{\mu\nu}
  = \osb{\Gamma}{LC}^\rho_{\mu\nu}
  + S^\rho{}_{\mu\nu}\,.
\end{equation}
From this, we obtained the following equation of motion for $S^\rho{}_{\mu\nu}$,
\begin{equation}
  \label{eq:einstein-palatini-shift-eom-app}
  0 =
  - g^{\mu\lambda} S^\nu{}_{\rho\lambda}
  - g^{\nu\lambda} S^\mu{}_{\lambda\rho}
  + \delta^\mu_\rho S^\nu{}_{\alpha\beta} g^{\alpha\beta}
  + g^{\mu\nu} S^\lambda{}_{\lambda\rho}\,.
\end{equation}
We can solve this using a similar strategy as we used for~\eqref{eq:einstein-palatini-direct-gamma-eom-app} above, and the computation is a bit more straightforward.
First, let us define the traces
\begin{equation}
  \label{eq:lor-S-traces}
  a_\mu
  =  S_{\mu\rho}{}^\rho\,,
  \qquad
  b_\mu
  = S^\rho{}_{\mu\rho}\,,
  \qquad
  c_\mu
  = S^\rho{}_{\rho\mu}\,.
\end{equation}
where we have reused some of the variable names in~\eqref{eq:app-direct-traces-def} to ease notation.
This allows us to rewrite the equation of motion~\eqref{eq:einstein-palatini-shift-eom-app} as
\begin{align}
  \label{eq:lor-S-eom}
  0 =
  - S^{\nu\rho\mu}
  - S^{\mu\nu\rho}
  + g^{\mu\rho} a^\nu
  + g^{\mu\nu} c^\rho\,.
\end{align}
Taking the trace over $\mu\rho$ and $\mu\nu$ in this equation, we find
\begin{equation}
  c_\rho = a_\rho\,,
  \qquad
  b_\rho = d a_\rho\,.
\end{equation}
With this, we can solve the equation of motion~\eqref{eq:lor-S-eom} by subtracting and adding two symmetric permutations, which leads to
\begin{equation}
  S^\rho{}_{\mu\nu}
  = \delta^\rho_\nu A_\mu\,,
\end{equation}
after setting $A_\mu = a_\mu$,
in agreement with Equation~\eqref{eq:einstein-palatini-shift-s-sol}.

\addcontentsline{toc}{section}{References}
\bibliographystyle{JHEP}
\bibliography{biblio}

\end{document}